\pdfoutput=1     
\documentclass[twocolumn]{aastex7}

\usepackage{graphicx}
\usepackage{amsmath}
\usepackage{booktabs}

\newcommand{\CMR}{C/MR^{2}}

\shorttitle{Ice Giants as Magma-Ocean Worlds}

\shortauthors{Young, Marcum, Werlen, \& Wulff}

\begin{document}

\title{\fontsize{14}{18}\selectfont Ice Giants Revisited: Uranus and Neptune as Magma Ocean Worlds}


\author[0000-0002-1299-0801]{Edward D. Young}
\affiliation{Department of Earth, Planetary, and Space Sciences, University of California, Los Angeles, CA 90095, USA}
\email[show]{eyoung@epss.ucla.edu}
\correspondingauthor{Edward D. Young}

\author[0009-0001-6858-9849]{Sarah P. Marcum}
\affiliation{Department of Earth, Planetary, and Space Sciences, University of California, Los Angeles, CA 90095, USA}
\email{smarcum13@ucla.edu}

\author[0009-0005-1133-7586]{Aaron Werlen}
\affiliation{Department of Earth, Planetary, and Space Sciences, University of California, Los Angeles, CA 90095, USA}
\email{werlen@ucla.edu}

\author[0000-0002-6788-8898]{Paula N. Wulff}
\affiliation{Department of Earth, Planetary, and Space Sciences, University of California, Los Angeles, CA 90095, USA}
\email{paulawulff@g.ucla.edu}

\begin{abstract}
Uranus and Neptune are commonly interpreted as volatile-rich ``ice giants'',
an assumption that underpins most interior models.  Here we show that their observed radii, bulk densities, gravitational harmonics, normalized moments of inertia, intrinsic luminosities, and key features of their atmospheric compositions are consistent with interiors comprising supercritical, hydrogen-rich magma oceans overlain by H$_2$-rich envelopes.  Our results, based on three fit parameters for each planet, provide a parsimonious explanation for the structures, thermal states, and atmospheric chemistries of Uranus and Neptune.  We find that the Solar System's ice giants can be understood as magma-ocean giants, with origins parallel to those of sub-Neptune gas-dwarf planets. A continuum among gas dwarf planets permits Neptune and Uranus to serve as accessible, data-driven test cases for structure models and material properties used to understand sub-Neptunes.
\end{abstract}


\section{Introduction}

Uranus and Neptune were originally classified as ``ice giants'' on
cosmochemical grounds.  For a solar-composition protoplanetary disk
beyond the snowline, the volatile compounds
H$_2$O, CH$_4$, and NH$_3$, comprising ``ices'', 
were predicted to outweigh refractory rock-forming material by roughly
a factor of two \citep{ReynoldsSummers1965, PodolakCameron1974,
PodolakReynolds1984, Podolak1995}, motivating the canonical three-layer
interior model of a rocky core, an icy mantle, and a hydrogen--helium
envelope \citep{HubbardMacFarlane1980, Marley1995}.

The canonical three-layer models for the interiors of Neptune and Uranus usually
include a rocky core, an intermediate ``icy'' mantle of water, methane, and
ammonia, and an outer H/He-rich envelope
\citep{HubbardMacFarlane1980, FortneyNettelmann2010, Nettelmann2013}.
However, data neither require the interiors of these ice giants to be
fully differentiated nor water-dominated in order to fit the observables.
Empirical structure models that fit the observed gravitational harmonics
without assuming a specific composition reveal that continuous density
gradients are equally consistent with the data and that rock-dominated
interiors cannot be ruled out
\citep{HelledAnderson2011, Helled2020b, Teanby2020, Neuenschwander2022, Neuenschwander2024}. 

The combination of a large number of model parameters and a comparatively
smaller number of observables has led to considerable degeneracies among
models for the structure and chemistry of the interiors of Uranus and
Neptune. We do not attempt to describe all successful models for the interiors of these planets, but we note that new variants that invoke larger fractions of rocky material compared with icy progenitors have been proposed recently \citep[e.g.,][]{Malamud2024, Tejada_Arevalo_2025, morfahelled2025, Ramirez2026}, as well as incorporation of exotic superionic phases \citep{Liu_2026_superionic}.  

In this paper we present modeling evidence that the ice giants of the Solar System
may actually be supercritical magma ocean giants.  The proposition is motivated in part by an evolving understanding of the astrophysical constraints on ices in protoplanetary disks and new findings regarding the physical chemistry of gas-dwarf planet interiors, both of which are in tension with the foundations of the canonical ice giant models.

Firstly, it is not clear that Uranus and Neptune formed from progenitors containing as much ice as is often cited. A budget of planet-forming materials based on solar elemental abundances shows that the commonly assumed ice-to-rock mass ratio of approximately 2:1 requires the condensation and incorporation of CO, and therefore formation beyond the CO snowline. Assigning the carbon in a solar-composition gas to CO, the oxygen remaining after CO formation first to silicates and then to water, and Fe to metal yields a total inventory composed of approximately 54 \% CO, 23 \% silicate, 12 \% water, and 11 \% Fe by mass. If the full CO inventory were incorporated into planetesimals, the condensed material beyond the CO snowline would therefore be approximately 66 \% ice by mass, consisting of 54 \% CO and 12 \% water. The dominance of CO ice persists even if one assumes that half of the carbon resides in organic solids. Between the water and CO snowlines, however, CO remains in the gas phase and does not contribute to the condensed planet-forming material. The remaining solids are then composed of approximately 50 \% silicate, 27 \% water, and 23 \% Fe by mass. Thus, water by itself comprises only roughly one quarter of the planet-forming material outside the water snowline but inside the CO snowline, excluding H$_2$ and He, considerably less than the 2:1 water-to-rock ratios assumed in many ice giant models. This calculation also has the merit of offering a simple explanation for the approximately 2:1 ratio of silicate to Fe metal in Earth. Timing of accretion, incomplete condensation, and the efficiency with which CO ice was incorporated into planetesimals affect the specific proportions, but an ice-dominated progenitor composition requires substantial accretion of CO-rich material from beyond the CO snowline.

The position of the CO snowline in the solar protoplanetary disk at any given time is poorly known, and observations of disks around low-mass pre-main-sequence stars allow for a wide range of possible locations. In TW Hya, often treated as a solar-nebula analog, N$_2$H$^+$ emission places the CO snowline near $\sim 30$ au, similar to the current position of Neptune but beyond that of Uranus, while rare CO isotopologue observations place the midplane CO snowline closer to $\sim 20$ au \citep{Qi2013,Schwarz2016,Zhang2017}, similar to the present-day semimajor axis of Uranus. Other T Tauri disks exhibit a similarly broad range of inferred CO snowline locations. CO isotopologue modeling yields midplane locations of $\sim 12\pm5$ to $\sim 30\pm5$ au for AS 209, IM Lup, and GM Aur \citep{Zhang2021}, whereas N$_2$H$^+$-based analyses suggest more distal CO and N$_2$ snowline positions. \citet{Qi_2019} report CO snowline radii of 58, 48, and 75 au for LkCa 15, GM Aur, and DM Tau, respectively, with corresponding N$_2$ snowline radii of 88, 78, and 145 au. These analogs indicate that, in solar-type protoplanetary disks during the T Tauri phase, the CO and N$_2$ snowlines may have resided beyond the present-day semimajor axes of Uranus and Neptune. Consequently, it cannot simply be assumed that the progenitors of either planet formed in a region where CO was available as a major condensed component.

The inference that the progenitors of the ice giants could have formed inside of the CO snowline is supported by the properties of the vestiges of planet formation in the outer Solar System. Kuiper Belt objects provide, arguably, the best direct evidence remaining of the composition of the outer planet-forming regions of the Solar System. These bodies are approximately 70 \% silicate \citep{Bierson2019}. The ices comprising Pluto are dominantly water, with CO, CH$_4$, and N$_2$ ices representing only a relatively thin veneer on the surface \citep{Stern2015,McKinnon2017,Schmitt2017}. Jupiter-family comets also sample volatile-rich planetesimals formed in the outer Solar System. Their compositions are consistent with those of Kuiper Belt objects in suggesting dust-to-ice ratios of about 3, corresponding to ice fractions of $\sim 25 \%$ \citep{Patzold2016,Fulle2019}.

The solar-abundance budget, astronomical constraints on snowline locations, and the compositions of surviving outer Solar System bodies therefore provide no compelling evidence that the progenitors of Uranus and Neptune must have been dominated by ice. Although condensation of CO could produce an ice-dominated inventory at sufficiently large heliocentric distances, observational evidence for bodies dominated by ices other than water remains limited. For these reasons, it is not at all clear that the progenitors of the interiors of planets formed in the outer Solar System contained the large ice fractions commonly assumed in ice giant models.

Secondly, chemistry at the conditions appropriate for the interiors of planets is equally important, if not more important, in determining the nature of their interiors than their radial position of accretion. This point is underscored by recent models for sub-Neptunes \citep{Schlichting_Young_2022, Young_2024, Young2025_Differentiation}. Copious amounts of water can be produced by reactions between primary hydrogen atmospheres and hot cores beneath  \citep{Kite2021,Schlichting_Young_2022, Young_Nature_2023, werlen_sub-neptunes_2025, Miozzi2025, Werlen2026}.  Therefore, the amount of H$_2$O in a planet is not directly related to where it accreted. Indeed, while more water may accrete to growing bodies in the outer protoplanetary disks, subsequent chemical evolution can lead to similar water concentrations for planets formed interior to the water ice line \citep{werlen_sub-neptunes_2025}.  What is more, the chemical memory of accreted water ices is, to first order,  most likely erased in the interiors of bodies due to miscibility with silicates. To see this, consider a planet assembled from planetesimals that were 25 \% by mass water ice, 50 \% rock (MgSiO$_3$), and 25 \% Fe (Earth-like rock/metal). DFT-MD calculations by \cite{Kovavcevic2022} indicate that molten MgSiO$_3$ and H$_2$O can become fully miscible at high pressure and temperature, similar to the behavior in the MgSiO$_3$-H$_2$ system. The initial 25 \% H$_2$O by mass contributes 22.2 \% oxygen and 2.8 \% hydrogen to the bulk planet. The oxygen will comprise part of the dense oxide-like component in the supercritical miscible phase, whereas the hydrogen increases the low-density component. Water-derived oxygen may be regarded as providing a means to oxidize Fe to FeO, increasing the rock/metal ratio, but this is an intra-phase redox redistribution and is not readily discernible in the fully miscible melt. The first-order effect of accreting water to an MgSiO$_3$ + Fe + H$_2$ mixture is therefore an increase in the effective hydrogen inventory, while the oxygen augments the dense interior. This water-derived H$_2$ is in addition to the free molecular H$_2$ accreted by the growing planet. Our models should largely capture this density effects in the form of total hydrogen.

Finally, the suggestion that the ice giants may be more rocky than previously thought \citep{Teanby2020} requires an analysis of the physical chemistry of ``rock" when intermingled with a large reservoir of H$_2$ at temperatures of thousands of degrees and pressures of tens to hundreds of GPa.  Recent work indicates that silicate, hydrogen, and iron are miscible at the conditions that prevailed in the interiors of ice giants, necessitating a reexamination of what is meant by ``rocky" in this context \citep{Young_2024, gilmore_core-envelope_2025, Young2025_Differentiation}.

This paper is organized as follows. In \S \ref{sec:constraints} we describe the observational constraints on the compositions of the ice giants.  In \S \ref{sec:model} we describe the supercritical magma ocean model, and in \S \ref{sec:methods} we describe our methods for deriving fits to the planet observables using this model. Results are described in \S \ref{sec:Results}, and various implications, predictions, and limitations are described in \S \ref{sec:discussion}.  We provide our conclusions in \S \ref{sec:conclusions}.

\section{Ice Giant Observational Constraints}
\label{sec:constraints}
The primary observational constraints used in interior structure
models for these planets are the 1-bar equatorial radius, bulk density, and
the gravitational harmonics $J_2$ and $J_4$.  The 1-bar radius and the
first determinations of $J_2$ and $J_4$ were established by the Voyager~2
flyby \citep{Tyler1989, Lindal1992} (Tables \ref{tab:neptune_constraints} and \ref{tab:uranus_constraints}).  No spacecraft has revisited either
planet since.  The $J_2$ and $J_4$ values used in modern interior models
are not the raw Voyager~2 numbers, however.  Rather, they have been
substantially refined through long-baseline astrometry of satellite orbital
motions \citep{Jacobson2009, Brozovic2020, Wang2023}. In addition, the gravitational parameters are composed of a static component due solely to oblateness and a dynamic component arising from winds \citep{Kaspi2013}. 

Published harmonics for Neptune are tabulated \citep{Wang2023} at a
reference radius $R_{\rm ref} = 25{,}225$~km \citep{Wang2023},
which exceeds the physical 1-bar equatorial radius
$R_{\rm 1bar}$ of $24{,}766$~km \citep{French2024} adopted as the
reference radius for our interior models.  We therefore renormalize the
$J_{2n}$ values to $R_{\rm 1bar}$,
\begin{equation}
J_{2n}(R_{\rm new}) = J_{2n}(R_{\rm ref})
    \left(\frac{R_{\rm ref}}{R_{\rm new}}\right)^{2n},
\label{eq:renorm}
\end{equation}
yielding $J_2 = 3528.91\times10^{-6}$ and $J_4 = -35.83\times10^{-6}$
at the 1-bar level, derived from the \citet{Wang2023} values.

Neptune's rotation period is fixed to the Voyager~2 magnetic field
periodicity of 58{,}000~s \citep{Tyler1989}, which is known to be
inconsistent with the planet's observed oblateness \citep{Helled2010,
Nettelmann2013}. This introduces a model systematic distinct from the
purely astrometric uncertainties in
Table~\ref{tab:neptune_constraints} \citep{Wang2023} . The normalized
moment of inertia $\CMR$ has not been directly measured for either planet,
but model values exist \citep{Nettelmann2013, BaileyStevenson2021,
Neuenschwander2022} that can serve as useful arbiters for structure models.

\begin{table*}[htbp]
\centering
\caption{Observational constraints used in the Neptune interior MCMC.
Gravitational harmonics from \citet{Wang2023} are renormalized from
$R_{\rm ref} = 25{,}225$ km to the 1-bar equatorial radius
$R_{\rm eq} = 24{,}766$ km via
$J_n(R_{\rm new}) = J_n(R_{\rm old})(R_{\rm old}/R_{\rm new})^n$,
increasing $J_2$ by $\sim$3.7\%. The $C/MR^2$ uncertainty reflects
the assumed model-to-observation systematic.}
\label{tab:neptune_constraints}
\begin{tabular*}{\textwidth}{@{}l@{\extracolsep{\fill}}cll@{}}
\hline\hline
Quantity & Value & Uncertainty & Source \\
\hline
Mass ($M$) &
  $1.02413 \times 10^{26}$ kg &
  fixed &
  \citet{Tyler1989} \\
1-bar equatorial radius ($R_{\rm eq}$) &
  $24{,}766$ km &
  $\pm 15$ &
  \citet{Lindal1992} \\
Rotation period ($P_{\rm rot}$) &
  $58{,}000$ s &
  fixed$^{a}$ &
  \citet{Tyler1989} \\
Equilibrium temperature ($T_{\rm eq}$) &
  $46.6$ K &
  $\pm 1.1$ K &
  \citet{PearlConrath1991} \\
\hline
\multicolumn{4}{@{}l}{Thermal constraints:} \\
1-bar temperature ($T_{1\,\rm bar}$) &
  $72$ K &
  $\pm 2$ K &
  \citet{Lindal1992} \\
Intrinsic luminosity ($L_{\rm int}$) &
  $3.3 \times 10^{15}$ W &
  $\pm\,0.35 \times 10^{15}$ W &
  \citet{PearlConrath1991} \\
\hline
\multicolumn{4}{@{}l}{Gravitational harmonics at
  $R_{\rm ref} = 25{,}225$ km:} \\
$J_2$ &
  $3401.655 \times 10^{-6}$ &
  $\pm\,3.994 \times 10^{-6}$ &
  \citet{Wang2023} \\
$J_4$ &
  $-33.294 \times 10^{-6}$ &
  $\pm\,10.000 \times 10^{-6}$ &
  \citet{Wang2023} \\
\hline
\multicolumn{4}{@{}l}{Renormalized to $R_{\rm eq} = 24{,}766$ km
  (this work):} \\
$J_2$ &
  $3528.912 \times 10^{-6}$ &
  $\pm\,4.143 \times 10^{-6}$ &
  \citet{Wang2023}, renorm. \\
$J_4$ &
  $-35.832 \times 10^{-6}$ &
  $\pm\,10.762 \times 10^{-6}$ &
  \citet{Wang2023}, renorm. \\
\hline
\multicolumn{4}{@{}l}{Model consistency target:} \\
Norm.\ moment of inertia ($C/MR^2$) &
  $0.2410$ &
  $\pm\,0.0048$ (2\%) &
  \citet{Nettelmann2013} \\
\hline
\multicolumn{4}{@{}l}{$^{a}$ Voyager~2 magnetic field periodicity;
  known to be inconsistent with Neptune's observed shape
  \citep{Nettelmann2013}. See text.} \\
\hline\hline
\end{tabular*}
\end{table*}

\begin{table*}[htbp]
\centering
\caption{Observational constraints used in the Uranus interior MCMC.
Gravitational harmonics from \citet{French2024} are reported at
$R_{\rm ref} = R_{\rm eq} = 25{,}559$ km. The $C/MR^2$ uncertainty reflects the
assumed model-to-observation systematic.}
\label{tab:uranus_constraints}
\begin{tabular*}{\textwidth}{@{}l@{\extracolsep{\fill}}cll@{}}
\hline\hline
Quantity & Value & Uncertainty & Source \\
\hline
Mass ($M$) &
  $8.68099 \times 10^{25}$ kg &
  fixed &
  \citet{JacobsonPark2025} \\
1-bar equatorial radius ($R_{\rm eq}$) &
  $25{,}559$ km &
  $\pm 4$ &
  \citet{Lindal1987} \\
Rotation period ($P_{\rm rot}$) &
  $62{,}064$ s &
  fixed &
  \citet{Desch1986} \\
Equilibrium temperature ($T_{\rm eq}$) &
  $58.1$ K &
  $\pm 1.0$ K &
  \citet{Pearl1990} \\
\hline
\multicolumn{4}{@{}l}{Thermal constraints:} \\
1-bar temperature ($T_{1\,\rm bar}$) &
  $76$ K &
  $\pm 2$ K &
  \citet{Lindal1987} \\
Intrinsic luminosity ($L_{\rm int}$) &
  $6.3 \times 10^{14}$ W &
  $\pm\,1.5 \times 10^{14}$ W &
  \citet{Wang2025} \\
\hline
\multicolumn{4}{@{}l}{Gravitational harmonics at
  $R_{\rm ref} = R_{\rm eq} = 25{,}559$ km:} \\
$J_2$ &
  $3509.291 \times 10^{-6}$ &
  $\pm\,0.412 \times 10^{-6}$ &
  \citet{French2024} \\
$J_4$ &
  $-35.522 \times 10^{-6}$ &
  $\pm\,0.466 \times 10^{-6}$ &
  \citet{French2024} \\
\hline
\multicolumn{4}{@{}l}{Model consistency target:} \\
Norm.\ moment of inertia ($C/MR^2$) &
  $0.2250$ &
  $\pm\,0.0045$ (2\%) &
  \citet{Nettelmann2013, PodolakHelled2012} \\
\hline\hline
\end{tabular*}
\end{table*}

The intrinsic luminosities of Uranus and Neptune provide additional
constraints on their interiors.  Intrinsic luminosities are inferred from
the planetary energy balance, requiring both the effective temperature
(from thermal emission) and the Bond albedo (from reflected sunlight),
the latter being the primary source of uncertainty.  The intrinsic
luminosities of Uranus and Neptune were first determined from
Voyager~2 IRIS data \citep{Pearl1990, PearlConrath1991}.  These data
indicate energy balances of $\mathcal{E} = 1.06 \pm 0.08$ and
$\mathcal{E} = 2.61 \pm 0.28$ respectively, where $\mathcal{E}$ is the
ratio of emitted to absorbed power related to the intrinsic luminosity by
$L_\mathrm{int} = (\mathcal{E}-1)(1-A)\pi R^2 \mathcal{F}_\odot$.  Here
$A$ is the Bond albedo, $R$ is the planet radius, and
$\mathcal{F}_\odot$ is the incident solar flux at the planet's orbital
distance.  The measured values yield intrinsic luminosities of
$L_\mathrm{int} \lesssim 3.5 \times 10^{14}$ W and
$3.3 \times 10^{15}$ W for Uranus and Neptune, respectively.  Neptune's
substantial internal heat source is well established, with albedo-driven
uncertainties of order 20\%, while Uranus' near-zero intrinsic luminosity
has been revised upward recently to $L_\mathrm{int} \approx
6.3 \times 10^{14}$ W \citep{Wang2025, Irwin2025}, driven primarily by a
reanalysis of the Bond albedo.  Compositional stratification and
limitations on convection have been invoked to match the observed thermal
states \citep{VazanHelled2020, Neuenschwander2024}.  Sources of
stratification proposed previously include phase separation between
partially miscible H$_2$ and H$_2$O \citep{BaileyStevenson2021,
CanoAmoros2024} and H$_2$O--CH$_4$--NH$_3$ exsolution to distinct water
and C--H--N-rich layers \citep{Militzer2024}.

The 1-bar temperatures, $T_{1\,{\rm bar}}$, provide additional constraints for interior models for both ice giants. Because the equilibrium temperatures are low, the values for $T_{1\,{\rm bar}}$ are sensitive to the deep structure that sets $L_{\rm int}$ as well as to the details of the opacities. 

The atmospheres of both planets are dominated by H$_2$ and He, with
CH$_4$ being the third-most abundant constituent.  In Neptune, the deep
tropospheric CH$_4$ mole fraction, retrieved below the
$\sim 1.5$~bar condensation level and taken to represent the well-mixed
bulk abundance, varies from 6 to 7\% at equatorial latitudes to $\sim 3\%$
near the poles \citep{Irwin2019muse, Irwin2021, Irwin2023muse,
Karkoschka2011}.  Uranus exhibits roughly half this abundance, with 1
to 4\% at low latitudes and $\sim 1\%$ near the poles below its
$\sim 1.2$~bar condensation level
\citep{Sromovsky2014, Sromovsky2019}.  In both cases the sub-condensation
mole fraction is the closest available proxy for the bulk atmospheric
carbon abundance, since deeper levels are inaccessible to remote
sensing.  These concentrations of methane imply carbon enrichments of
$\sim 80\times$ solar for Neptune's atmosphere and $\sim 40$ to $50\times$
solar for Uranus' atmosphere \citep{Cavalie2020}.  H$_2$S has also been
detected near 1 to 4 bar in both planets with mole fractions of 0.4 to
0.8~ppm for Uranus and $\sim 1$ to 3~ppm for Neptune.  NH$_3$ is
undetected at any level in both atmospheres \citep{dePater1991}, although its presence has been inferred on the basis of latitudinal variations in microwave opacities \citep{Hofstadter2003}.  The
detection of gaseous H$_2$S and the non-detection of NH$_3$ implies
S/N substantially greater than solar in the accessible regions of the
atmospheres for both planets \citep{Irwin2018h2s, Irwin2019h2s,
dePater1991}.  Stratospheric photochemistry driven by CH$_4$ photolysis
produces C$_2$H$_6$ and C$_2$H$_2$, detected in both atmospheres,
together with trace CO and HCN \citep{Moses2020}.

\section{Model}
\label{sec:model}

A growing understanding of the physical chemistry of the
silicate--hydrogen--iron system is elucidating the nature of gas dwarf
sub-Neptune planets.  We use the term gas dwarf here to refer to planets
thought to consist of magma oceans\footnote{Because the supercritical mix is so hydrogen rich, the use of the term ``magma" may no longer be entirely appropriate. We use it to distinguish it from the even more H$_2$-rich envelope. } overlain by H$_2$-rich envelopes but
small enough to have avoided runaway growth.  We have learned that at
the temperatures and pressures imposed by accretion and self gravity,
the interiors of these planets are likely to be a single, supercritical
mixture of silicate, iron, and hydrogen, modeled as miscible MgSiO$_3$,
Fe, and H$_2$ based on ab-initio molecular dynamics (DFT-MD) simulations
\citep{gilmore_core-envelope_2025, Young_2024,
Young2025_Differentiation, RogersYoungSchlichting2025_MNRAS}.  These
calculations are supported by recent experimental findings
\citep{Miozzi2025}.  The boundary between this supercritical magma ocean
and an overlying H$_2$-rich envelope is a first-order phase transition
with coexisting compositions in the MgSiO$_3$--H$_2$ system defined by a
pressure-dependent binodal in temperature ($T$) versus composition
($x_{\rm H_2}$) space.  The same DFT-MD simulations that inform us about
the phase equilibria also provide constraints on the equation of state
(EoS) for the supercritical magma oceans.

\begin{figure*}[!t]
\centering
\includegraphics[width=0.7\textwidth]{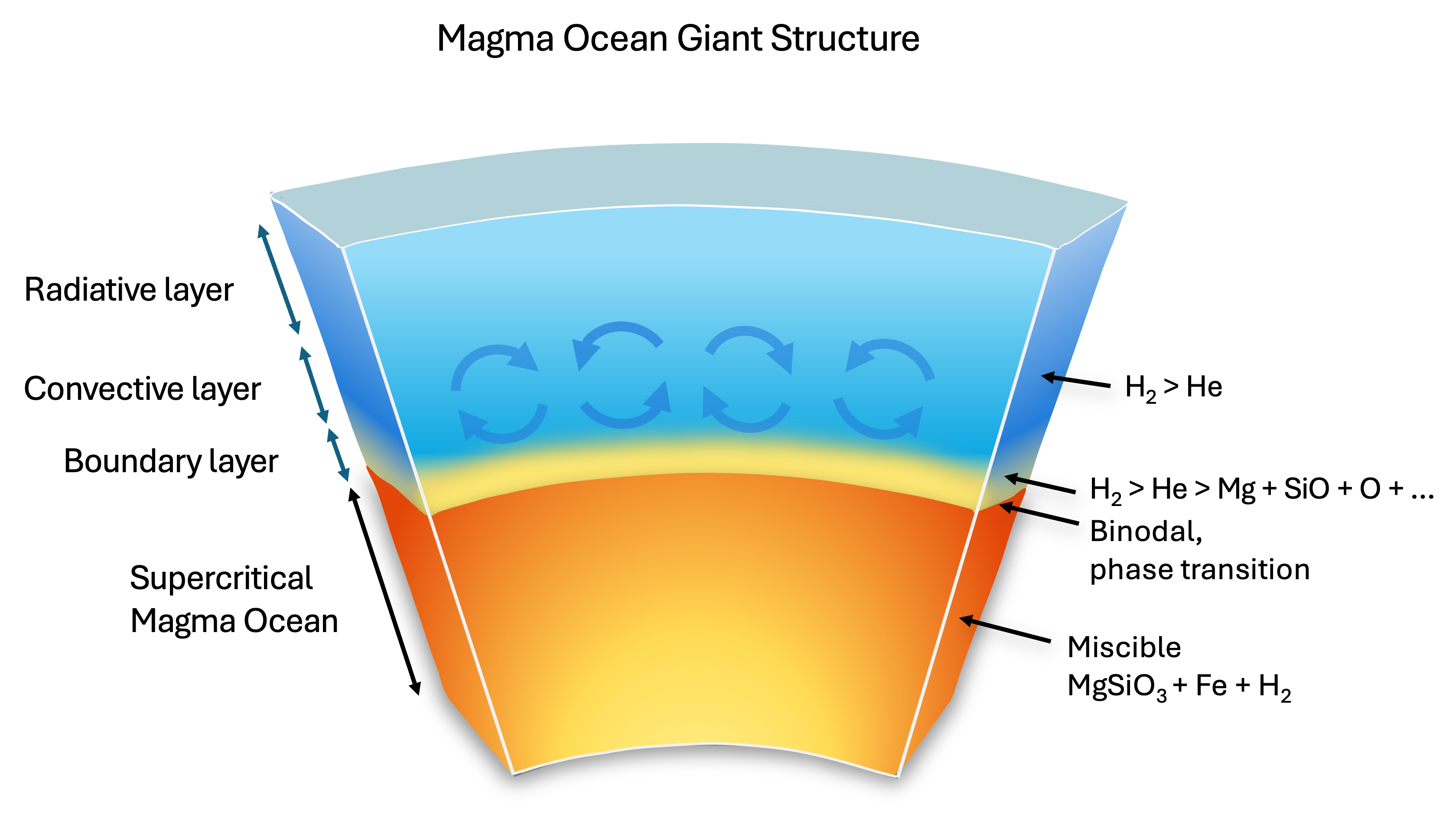}
\caption{Schematic showing the layered structure of Neptune and Uranus
         as proposed here (not to scale).}
\label{fig:schematic}
\end{figure*}

The picture of these planets that emerges from the new findings is one
composed of four main layers (Figure~\ref{fig:schematic}).  The deepest
layer is a magma ocean composed of a supercritical mixture of silicate,
iron, and hydrogen, similar to mixtures of MgSiO$_3$, Fe, and H$_2$.
For weight percent fractions of total hydrogen, the density of the
overlying hydrogen-rich envelope is sufficient to preserve a molten magma
ocean for several Gyr timescales \citep{ginzburg2016a}.  At large radial
distances from the center, at the conditions specified by the binodal for
the first-order phase transition, the magma ocean gives way to an
H$_2$-rich envelope phase.  At the base of the envelope is a boundary
layer in which convection is inhibited \citep{Ahlers_2009}, augmented by
molecular weight gradients predicted by the phase equilibria
\citep{Young_2024, Markham2022, Misener2023}.  Estimates for the
Rayleigh number in the supercritical melt are sufficiently high that the
supercritical magma ocean should be actively convecting 
\citep{Young_2024}.  Above the basal boundary layer, the envelope is
convective up to altitudes where radiative diffusion dominates over
convective heat transport.

Details of the planet model, including material properties, equations of
state, phase equilibria, and governing equations are given in
Appendices A through D.  Summarizing the salient features briefly
here, the supercritical magma ocean has a density much lower than that
of pure MgSiO$_3$. Differences in the pressure-dependence of $\rho$ from 
pure MgSiO$_3$ is constrained from DFT-MD simulations at several hydrogen concentrations (see Appendix D).  Hydrogen alters the compression behavior of the liquid. While the reference bulk modulus, $K_T$, is comparable in the equations of state of pure MgSiO$_3$ and the supercritical mixture of MgSiO$_3$ and $4\%$ H$_2$ by mass, the pressure derivative of the bulk modulus decreases from 5.71 for dry MgSiO$_3$ liquid to 4.32 for the H-bearing mixture \citep{Marcum2026}. This lower $K’$ implies that the supercritical mixture stiffens less rapidly with increasing pressure. Thus, hydrogen results in a lower density and makes the phase more compressible at high pressure. We use a linear fit of $K'$ with hydrogen concentration to extrapolate the EoS to higher concentrations of H$_2$ (Appendix D).

The fluid comprising the
magma ocean phase is expected to have a melting temperature of order
1000 degrees lower than that of pure MgSiO$_3$ due to the melting-point
depression effect of mixing (Appendix D).  Iron that
would otherwise form an iron metal core is entirely miscible in the
MgSiO$_3$--H$_2$ melt, based on phase equilibria in the ternary
MgSiO$_3$--Fe--H$_2$ system \citep{Young2025_Differentiation}, and has a limited effect on the density (Appendix D).
The physical chemistry thus ensures that the interior of the planets is
fully molten and composed of a single convecting phase.  Immediately
above the magma ocean, the binodal specifies that there is a heavy load
of Mg, Si, and O present in the hydrogen-dominated phase.  We consider
that the silicate species will condense moving outward as temperature
decreases.  We account for condensation as a Rayleigh-like distillation
as silicates and oxides rain out back to the surface.  This leads to a
significant molecular weight gradient that stabilizes the base of the
envelope against convection.  The Rayleigh number relative to the
critical Rayleigh number, $Ra/Ra_{\rm crit}$, is $\ge 10^9$, with the
determination of precise values hampered by uncertainties associated
with various scaling laws.  Nonetheless, once the molecular weight
gradient subsides, the envelope should be actively convecting.  We
employ a moist adiabat to account for the enthalpy effect of
condensation.  The convective layer transitions smoothly into a
radiative diffusion regime above, and eventually to a radiative layer,
which we model using a two-stream solution
(Appendix A).

\section{Methods} 
\label{sec:methods}
We have applied this picture of gas-dwarf planets to Uranus and Neptune.
We modeled the interior structures of Uranus and Neptune with
\texttt{Planet\_LAB}\footnote{\url{https://github.com/eyoungucla/planet_Lab}; https://doi.org/10.5281/zenodo.22240097},
a forward code that constructs a one-dimensional spherical planet
realization from three parameters.  The parameters
are the pressure of the binodal separating the magma ocean from the
overlying envelope, $P_{\rm binodal}$, for the planet, the total
H$_2$ mass fraction for the planet, $x_{\rm H_2}$, and the ratio of the Rayleigh number for the atmosphere to the critical Rayleigh number that specifies the width of the boundary layer at the base of the atmosphere, $Ra/Ra_{\rm crit}$ (Appendices~A--D).
For each proposed realization, \texttt{Planet\_LAB} integrates
hydrostatic equilibrium on spherical shells outward, crossing through a
binodal-defined magma ocean-envelope boundary. We use trapezoidal quadrature
over $1.2\times10^5$ radial steps, returning the density profile
$\rho(r)$ and the radius $R_{1\,\rm bar}$ at which the column
reaches 1\,bar. Because the integration is spherical, $R_{1\,\rm bar}$
is a volumetric mean radius. The mass and (spherical) moment of
inertia are
\begin{align}
M &= \int_0^{R_{1\,\rm bar}} 4\pi r^2\,\rho(r)\,dr, \\
I &= \int_0^{R_{1\,\rm bar}} \tfrac{8\pi}{3}\,\rho(r)\,r^4\,dr,
\end{align}
with the normalized moment of inertia $C/MR_{1\,\rm bar}^2 = I/(MR_{1\,\rm bar}^2)$.

The gravitational moments and rotational figures are computed a posteriori from the spherical density profiles using the concentric Maclaurin spheroid (CMS) method \citep{Hubbard2013},
implemented in the \textsc{Python} translation of the original
\textsc{Matlab} code \citep{Movshovitz2019}. We pass $\rho(r)$,
$R_{1\,\rm bar}$, and the published rotation period to the CMS module, which
iterates the equipotential surfaces of the rotating fluid figure and
returns the dimensionless multipole coefficients $J_2$ and $J_4$.
Fully self-consistent oblate methods --- the Theory of Figures
\citep{Nettelmann2013, HelledFortney2020} and the natively-coupled
CMS treatments of \citet{Wahl2017} and \citet{Movshovitz2019} ---
iterate the density profile and the equilibrium shape simultaneously
on level surfaces, allowing the density distribution to feed back
into the rotating figure as it relaxes. The CMS as implemented here solves for the
shape and gravity moments self-consistently at the published rotation
rate but treats the input $\rho(r)$ as fixed. Allowing the density
profile to relax in response to rotational deformation would modify
$J_n$ at the $\mathcal{O}(q_{\rm rot}^2)$ level where
$q_{\rm rot} = \Omega^2 R^3 / GM \approx 0.026$ for both planets
\citep{Schubert2004}; the shifts are less than the observational uncertainties and
certainly less than systematic uncertainties related to the dynamic contributions. 
The use of a 1D structure code is motivated by the miscibility/immiscibility
phase boundary. The radial position of the H$_2$-silicate binodal is determined by local
$(P, T)$ conditions, and locating it self-consistently across an
MCMC ensemble requires evaluating that boundary at every realization.
A 1D spherical model permits mapping the boundary as a single radius at
each step. Because exsolution creates compositionally stratified layers in which
the isothermal level-surface assumption fails, a self-consistent oblate method would require coupling
the figure iteration to a 2D heat-transport treatment, which is computationally expensive and beyond the present scope.  

In the CMS calculation, lengths are normalized internally by an outer
spheroid radius $a_0$. Because CMS as implemented here computes the gravity field of a
prescribed density distribution rather than iterating the shape and
structure jointly, the choice of $a_0$ represents the reference radius
for the $J_n$ values. We adopt
$a_0 = R_{1\,\rm bar}$ so that CMS produces $J_n$ at the volumetric
reference radius determined by our spherical structure integration; the
$J_n$ are then renormalized to the equatorial reference (below) permitting 
comparisons to published values. For this purpose we apply a first-order
oblate-spheroid bulge correction, where the flattening $f$ is
\begin{equation}
f \;\approx\; \frac{3 J_2 + q_{\rm rot}}{2},
\label{eq:flattening}
\end{equation}
\noindent and apply it using
\begin{equation}
R_{\rm eq,\,model} \;=\; R_{1\,\rm bar}\,(1 + f/3)
\label{eq:bulge}
\end{equation}
such that 
$J_n \to J_n (R_{1\,\rm bar}/R_{\rm eq,\,model})^n$. The first-order
truncation in Equation~\ref{eq:bulge} is accurate to
$\mathcal{O}(f^2) \approx 3\times 10^{-4}$, well below all reported
uncertainties.

From the planet model realizations, including the shape models, we
obtained solutions that minimize the residuals between the model and
six observable or derived quantities for Neptune and Uranus: the
gravitational harmonics $J_2$ and $J_4$ (compared to wind-corrected
values described below), the normalized moment of inertia $C/MR^2$, the 1-bar
equatorial radius $R_{\rm eq}$, the 1-bar
temperature $T_{1\,\rm bar}$, and the intrinsic luminosity
$L_{\rm int}$.

We modify our target $J_2$ and $J_4$ values for our static models by removing the dynamic component. For this purpose we perform a manual iteration between our planet realizations and models for the winds (Appendix E), the latter following methods described by  \cite{Dietrich_2021} and adapted to the Ice Giants as in \citet{Soyuer_2023}.

Most previous works linking Uranus' and Neptune's zonal winds to dynamic gravity contributions \citep[e.g.][]{Kaspi2013} assume that $J^{\rm obs}_2 = J^{\rm stat}_2$, so that only $J_4$ constrains the wind-induced component of the gravity signature. By assuming $J^{\rm dyn}_2 = 0$, deep-reaching winds are excluded \textit{a priori}.  However, deeper winds necessarily generate a non-negligible $J^{\rm dyn}_2$. As the problem is degenerate with only two measured gravity moments, we assume a decay function of the winds with depth that is physically motivated, and allow them to extend to the bottom of the convective atmosphere. Details of this calculation and the $J_2^{\rm dyn}$ and $J_4^{\rm dyn}$ values associated with the final models for Uranus and Neptune are presented in Appendix E.

In the results presented here, we find that for Uranus, winds extending down to the base of the convective layer, a depth of 4584 km,  $\Delta J_2 = J^{\rm obs}_2 - J^{\rm stat}_2$ $= -13.47\times 10^{-6}$  and $\Delta J_4=-3.46\times 10^{-6}$. For Neptune, winds extending to the base of the convective layer at a depth of 
3838 km, yield $\Delta J_2  = 0.87\times 10^{-6}$  and $\Delta J_4=-3.88\times 10^{-6}$ (Appendix E). 
The dynamic contribution to $J_2$ is negative for Uranus but positive for Neptune, while both dynamic contributions to $J_4$ are negative. The correction to $J_2$ for Uranus is $\sim 30$ times the quoted measurement uncertainty (Table \ref{tab:uranus_constraints}) while for Neptune it is $0.2$ times the quoted uncertainty (Table \ref{tab:neptune_constraints}).  The corrections for $J_4$ for Uranus and Neptune are 7 and $0.4$ times nominal uncertainties. The smaller  correction/uncertainty ratios for Neptune reflect the Monte Carlo error analysis for observed values by \cite{Wang2023}. 

The low equilibrium temperatures for the ice giants results in the 1-bar temperature being relatively insensitive to the shortwave irradiation
term. In the grey atmosphere calculation, variations in
\(\gamma=\kappa_{\rm vis}/\kappa_{\rm IR}\) have little effect on
\(T_{1\,{\rm bar}}\), and the 1-bar level is more closely
coupled to the intrinsic thermal structure imposed by the intrinsic luminosity, the pressure--optical-depth
relation, and the radiative--convective transition.  We therefore include
\(T_{1\,{\rm bar}}\) as a constraint on the coupled
interior--atmosphere model, while assigning it an estimated systematic 
model uncertainty of $\pm 5$ degrees, larger than the formal observational error but commensurate with probable uncertainties in opacities (see \S \ref{sec:discussion}). Previous models have generally either anchored an assumed adiabat at the observed $T_{\rm 1\,bar}$ \citep[e.g.,][]{BaileyStevenson2021, Nettelmann2013}, adjusted the profile by a scale factor to match observations \citep[e.g.,][]{Nettelmann2016}, or omitted this constraint entirely where the interior is the sole focus. These options are not justified where the lower boundary conditions for the atmospheres are constrained, in this case by the binodal phase boundary. 

The best fit is obtained using  Markov Chain Monte Carlo (MCMC) searches applied to the forward planet model code that explores the 3-D parameter space to find the posterior distribution over $(P_{\rm binodal},\, x_{\rm H_2}$, $Ra/Ra_{\rm crit})$.  These three parameters are used to obtain the maximum likelihood fit for the six observables. 

\section{Results}
\label{sec:Results}

\begin{figure*}[!t]
\centering
\includegraphics[width=\textwidth]{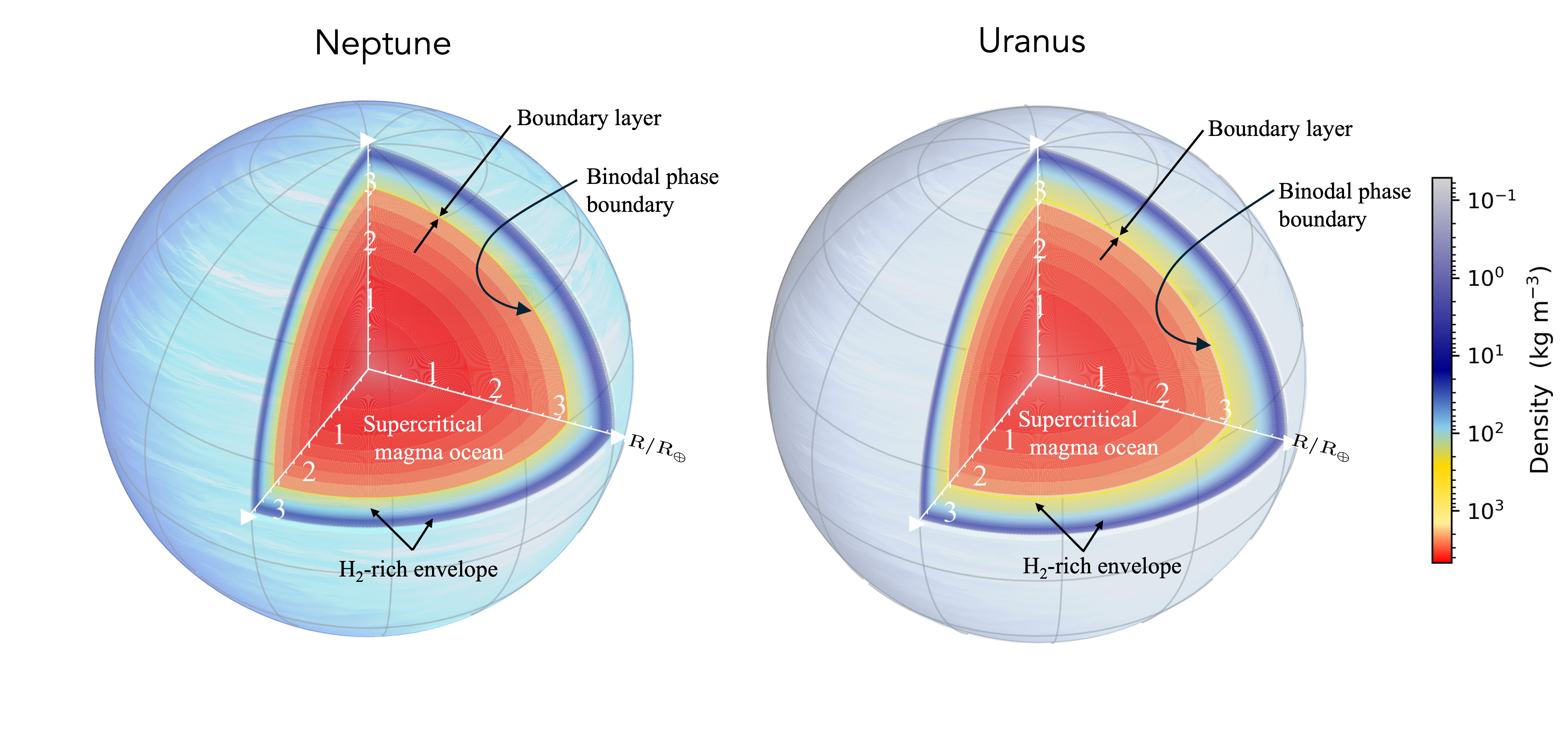}
\caption{Models for Neptune and Uranus derived in this study expressed as density versus radius.
Contour colors show density. Key elements of the models are labelled. The bright narrow band at the base of the
envelope in each case is the boundary region stable against convection, $\sim 120$ km and $\sim 180$ km wide in the cases of Neptune and Uranus, respectively. }
\label{fig:Neptune_and_Uranus}
\end{figure*}

\subsection{Neptune}

Our best-fit model parameters for Neptune are summarized in Table \ref{tab:neptune_bestfit}. The best-fit model expressed in density versus radius and various features of the model are shown in Figure \ref{fig:Neptune_and_Uranus}. 
The fixed input parameters are the total mass of $17.150\,M_\oplus$ and the
equilibrium temperature of 47~K. 
By adjusting the binodal pressure (the pressure at the phase transition between the
magma ocean and the overlying envelope), the total hydrogen fraction of the planet by weight, and $Ra/Ra_{\rm crit}$ we fit the equatorial 1-bar radius,  external gravitational harmonics coefficients, adjusted for dynamic effects, inferred normalized moment of inertia, 1-bar temperature, and intrinsic luminosity with a reduced $\chi^2$ of $2.6$, based on three degrees of freedom and the uncertainties shown in (Table \ref{tab:neptune_constraints}). This corresponds to an upper-tail p value of $\simeq 0.05$.  The reduced $\chi^2$ value $>1$ is driven entirely by a 12 K overestimate of the 1-bar temperature (Table \ref{tab:neptune_bestfit}).

The uncertainties used to calculate the reduced $\chi^2$ comprise a mixture of observational and model uncertainties, whereby purely observational values for $\sigma$ are used  for three of the variables and systematic errors associated with modeling are used for the other three.  The uncertainties on $J_2$ and $L_{\rm int}$ are purely observational, taken directly from the published values. We could also use $|\Delta J_2|$, the absolute value of our dynamic correction to $J_2$,  as a potential systematic uncertainty, but for Neptune the wind correction to $J_2$ is smaller than the observational uncertainty itself ($|\Delta J_2|/\sigma_{J_2,\rm obs} \approx 0.2$), so no broadening was necessary. The uncertainty on $R_{\rm eq}$ is also observational, where the adopted $\sigma_R$ of $22$~km is intermediate between the reported uncertainties on the equatorial and polar 1-bar radii of $\pm 15$~km and $\pm 30$~km respectively, both
dominated by atmospheric rotation-period scatter \citep{Lindal1992}.  The uncertainties on $J_4$, $T_{\rm 1\,bar}$, and $C/MR^2$ include known systematic model errors that are greater than the purely observational values. For $J_4$,
we adopt $\sigma_{J_4} = |\Delta J_4|$, approximately half the
Monte-Carlo uncertainty of \citet{Wang2023}
(Table~\ref{tab:neptune_constraints}), where $\Delta J_4$ is
our calculated wind-correction to the static value.  For $T_{\rm 1\,bar}$, we
adopt $\pm 5$~K rather than Lindal's observational $\pm 2$~K, as a
concession to systematic uncertainties in atmospheric opacities
$\kappa$ that propagate into the modeled temperature at the 1-bar
level \citep{Fletcher2018, Siebenaler2026}.  For $C/MR^2$, we adopt a
2\% model uncertainty \citep{MovshovitzFortney2022, Siebenaler2026}. The best-fit solution fits all parameters to within the nominal observational uncertainties, excluding $T_{\rm 1\, bar}$, even where these limits were relaxed to account for model systematic errors in the search (compare Tables \ref{tab:neptune_bestfit} and \ref{tab:neptune_constraints}). 

We also ran searches that matched the static (uncorrected) $J_2$ and
$J_4$ values, using the larger Monte-Carlo-based $J_4$ uncertainty of
\citet{Wang2023} (not reported here).  In all cases, the best-fit model $J_4$ approached the wind-corrected value, about $4 \times 10^{-6}$ greater (smaller absolute value) than the static observed value, in agreement with our independently
calculated wind effect.

\begin{figure*}[!t]
\centering
\includegraphics[width=\textwidth]{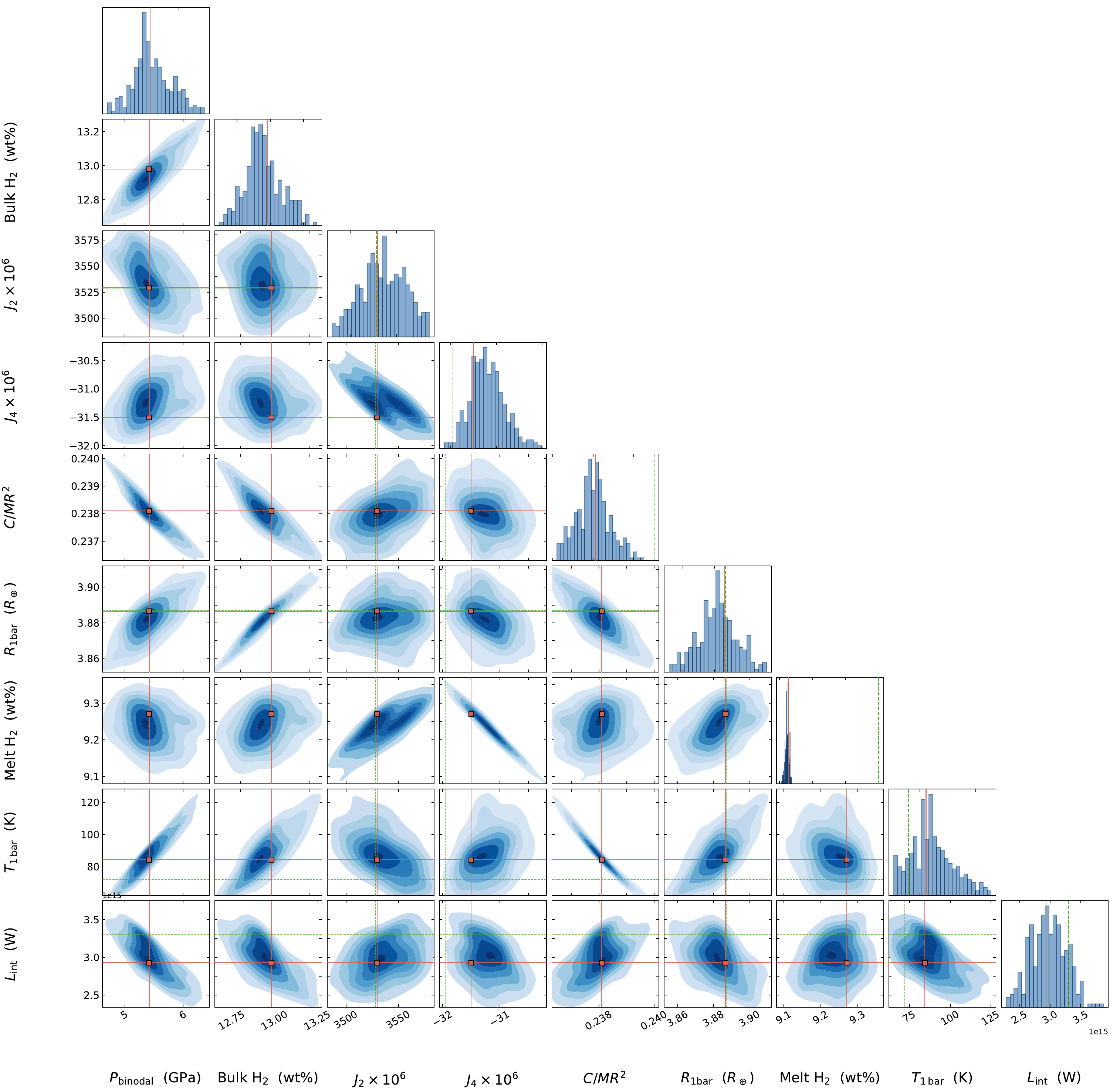}
\caption{Corner plot for MCMC search over 1,100 models for the best-fit for Neptune showing the marginalized posterior densities for the parameters of interest. The melt H$_2$ weight fraction that was not part of the fit is shown for reference. Green dashed lines are target values, and the red squares and red lines show the highest likelihood solution (line for melt H$_2$ is maximum allowed in search).  }
\label{fig:Neptune_corner}
\end{figure*}

The maximum likelihood solution from the MCMC search is obtained for $P_{\rm binodal} =  5.42$~GPa and a bulk planet H$_2$ fraction of $12.98\%$ by mass. The overall best-fit structure is composed of a supercritical magma ocean extending out to $3.15 R_{\oplus}$ and an H$_2$-rich envelope extending out to a 1-bar equatorial radius of $3.863 R_{\oplus}$ (Figure \ref{fig:Neptune_and_Uranus}).  These best-fit binodal pressure and hydrogen fraction values are strongly positively correlated, as shown in the search corner plot in Figure \ref{fig:Neptune_corner}.  We note that reasonable changes to the estimates for the equation of state for the supercritical magma ocean (e.g., behavior of compressibility with hydrogen content)  can change the precise values for these parameters without meaningfully degrading the quality of the fit; the magma ocean adiabat derived by the MCMC search is effectively the same, but with modestly different values for $x_{H_2}$ and $P_{\rm binodal}$.

\begin{table}
\centering
\caption{Best-fit solution from the Neptune interior MCMC.
The reduced $\chi^2$ is calculated with $n_{\rm obs} = 6$ constraints
and $n_{\rm free} = 3$ free parameters. A one-sided prior on the
melt-layer H$_2$ content (ceiling $12.0$~wt\%, $\sigma = 1.0$~wt\%)
is included but inactive at the best fit (predicted $9.27$~wt\%).}
\label{tab:neptune_bestfit}
\begin{tabular*}{\columnwidth}{@{\extracolsep{\fill}}lccc@{}}
\hline\hline
Quantity & Predicted & Target & $\chi^2$ \\
\hline
$P_{\rm binodal}$ (GPa)   & 5.42                 & ---                                  & --- \\
Bulk H$_2$ (wt\%)         & 12.98                & ---                                  & --- \\
$Ra/Ra_{\rm crit}$             & $1.224\times10^{12}$  & ---                             & --- \\
\hline
$J_2$ ($\times 10^{-6}$, static) & 3529.50       & $3528.05 \pm 4.140$                 & 0.122 \\
$J_4$ ($\times 10^{-6}$, static) & $-31.500$     & $-31.952 \pm 3.500$                 & 0.017 \\
$C/MR^2$                         & 0.2381        & $0.2410 \pm 0.00482$                & 0.362 \\
$R_{\rm eq}$ (km)                & 24760.9       & $24766.0 \pm 22.0$                  & 0.054 \\
$T_{\rm 1\,bar}$ (K)             & 84.4          & $72.0 \pm 5.0$                      & 6.150 \\
$L_{\rm int}$ (W)                & $2.93\times10^{15}$ & $3.30 \pm 0.35\times10^{15}$ & 1.118 \\
\hline
\multicolumn{3}{@{}l}{$\chi^2_{\rm total}$}                                              & 7.823 \\
\multicolumn{3}{@{}l}{$\chi^2_{\rm red}$ ($\nu = 3$)}                                    & 2.608 \\
\hline\hline
\end{tabular*}
\end{table}

 Interestingly, $J_4$, a parameter most sensitive to the outer structure of the planet, exhibits a strong negative correlation with the H$_2$ concentration in the supercritical melt phase (Figure \ref{fig:Neptune_corner}). This is an illustration of  the integrated nature of the structure model.  It can be traced to the links between $P_{\rm binodal}$ and $T$ at the base of the hydrogen-rich envelope, and the concentration of H$_2$ in melt imposed by these conditions at the binodal boundary. 

The various model parameters that characterize the model Neptune are listed in Table \ref{tab:nep_ura_compare}.  Salient features of the model include the binodal temperature of 3256 K, corresponding to the boundary between the supercritical magma ocean and the H$_2$-rich envelope, an envelope mass fraction for the planet of $0.04$ (note in our model we used a molecular weight of $2.3$ g/mol to accommodate a nominal He fraction), and a Ledoux-stable boundary layer that is 120 km thick above the magma ocean. The model intrinsic luminosity of $2.9\times 10^{15}$ W is virtually indistinguishable from the measured value for Neptune of $3.3 \pm 0.7 \times 10^{15} $ W for the nominal 1-bar equatorial radius \citep{PearlConrath1991}. 

\begin{figure}[t!]
  \centering
  \includegraphics[width=0.99\columnwidth]{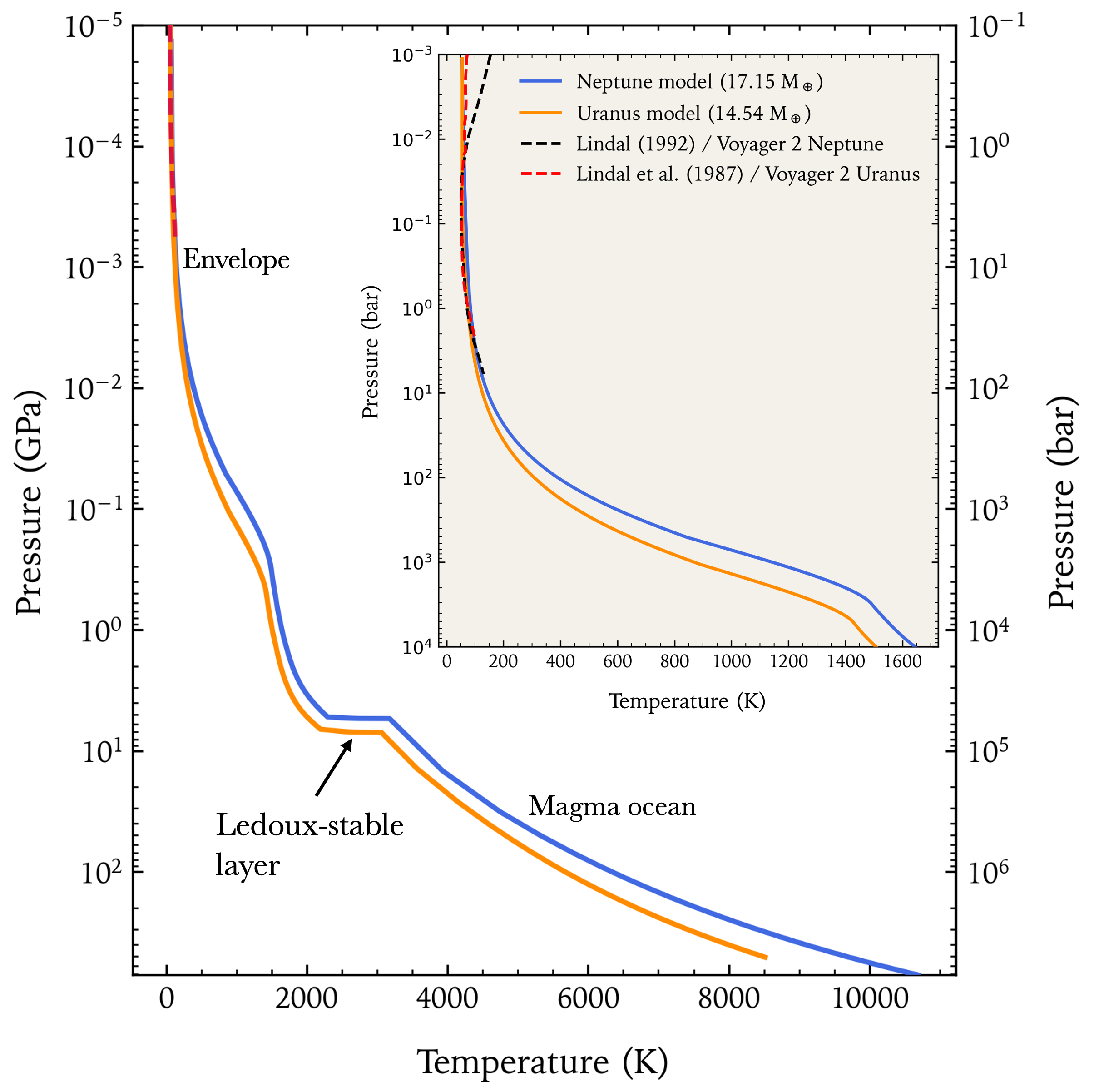}
  \caption{Model temperature-pressure profiles for Neptune and Uranus derived in this study. The regions corresponding to the supercritical silicate-hydrogen-iron magma ocean and the overlying Ledoux-stable layer (boundary layer) in the hydrogen-rich envelope are indicated. The inset shows the upper atmosphere compared with the Voyager 2 observations for both planets.}
  \label{fig:TandP}
\end{figure}

The temperature--pressure profile for the model Neptune is shown in Figure~\ref{fig:TandP}. The effect of the Ledoux-stable boundary layer on temperature is clearly evident. Our 1-bar temperature for Neptune is about 12 degrees greater than the measured value of $72\pm2$\,K \citep{Lindal1992}. 

\begin{table*}[t]
\centering
\caption{Comparison of best-fit interior model outputs for Neptune and Uranus.}
\label{tab:nep_ura_compare}
\begin{tabular*}{\textwidth}{@{\extracolsep{\fill}}lcc@{}}
\hline\hline
Quantity & Neptune & Uranus \\
\hline
\multicolumn{3}{@{}l}{\textit{Bulk planet}} \\
Mass (M$_\oplus$)                       & 17.150 & 14.536 \\
Radius at 1\,bar (R$_\oplus$)           & 3.863  & 3.986 \\
Core mass (M$_\oplus$)                  & 16.449 & 13.894 \\
Core radius (R$_\oplus$)                & 3.151  & 3.114 \\
Core volume fraction                    & 0.537  & 0.472 \\
Atm.\ volume fraction (chord)           & 0.463  & 0.528 \\
\hline
\multicolumn{3}{@{}l}{\textit{Composition (H$_2$ mass fraction)}} \\
Whole planet                            & 0.1298 & 0.1436 \\
Core                                    & 0.0927 & 0.1041 \\
Gas envelope (H$_2$ + He, $\mu = 2.3$)  & 0.9994 & 0.9981 \\
Silicate condensates                    & 0.0201 & 0.0173 \\
Gas mass fraction of planet             & 0.0393 & 0.0424 \\
Gas mass percent of planet (\%)         & 3.927  & 4.235 \\
\hline
\multicolumn{3}{@{}l}{\textit{Atmosphere / envelope}} \\
Surface pressure (GPa)                  & 5.420  & 7.040 \\
Surface temperature (K)                 & 3256.3 & 3071.5 \\
Equilibrium temperature (K)             & 47.0   & 58.1 \\
Mass of atmosphere (kg)                 & $4.186\times10^{24}$ & $3.839\times10^{24}$ \\
Atm.\ density, $R$ $ < 1\,\mathrm{bar}$ (kg\,m$^{-3}$) & 146.68 & 106.98 \\
Atm.\ volume, $R$ $ < 1\,\mathrm{bar}$ (m$^3$)         & $2.854\times10^{22}$ & $3.588\times10^{22}$ \\
Pressure at $R_{\rm rcb}$ (bar)         & 5  & 7 \\
$R(1\,\mathrm{bar})/R_{\rm c}$          & 1.226  & 1.280 \\
$Ra/Ra_{\rm crit}$\ (envelope)             & $1.224\times10^{12}$ & $5.674\times10^{9}$ \\
Boundary layer thickness (m)            & 105.4  & 741.1 \\
Ledoux layer above surface (m)          & $1.219\times10^{5}$ & $1.842\times10^{5}$ \\
Internal luminosity $L_{\rm int}$ (W)   & $2.933\times10^{15}$ & $6.646\times10^{14}$ \\
\hline
\multicolumn{3}{@{}l}{\textit{Core}} \\
Central temperature (K)                 & 10698.1 & 8516.8 \\
Bulk core density (kg\,m$^{-3}$)        & 2899.4  & 2537.3 \\
Core thermal energy (J)                 & $1.240\times10^{33}$  & $8.617\times10^{32}$ \\
Core grav.\ potential energy (J)        & $-2.063\times10^{34}$ & $-1.474\times10^{34}$ \\
\hline
\multicolumn{3}{@{}l}{\textit{Energy budget}} \\
Latent heat to form atmosphere (J)      & $9.665\times10^{31}$  & $8.102\times10^{31}$ \\
Planet thermal energy (J)               & $1.276\times10^{33}$  & $9.135\times10^{32}$ \\
Planet grav.\ potential energy (J)      & $-2.129\times10^{34}$ & $-1.561\times10^{34}$ \\
Total energy (J)                        & $-2.010\times10^{34}$ & $-1.444\times10^{34}$ \\
\hline\hline
\end{tabular*}
\end{table*}

\begin{figure}[t!]
  \centering
  \includegraphics[width=0.99\columnwidth]{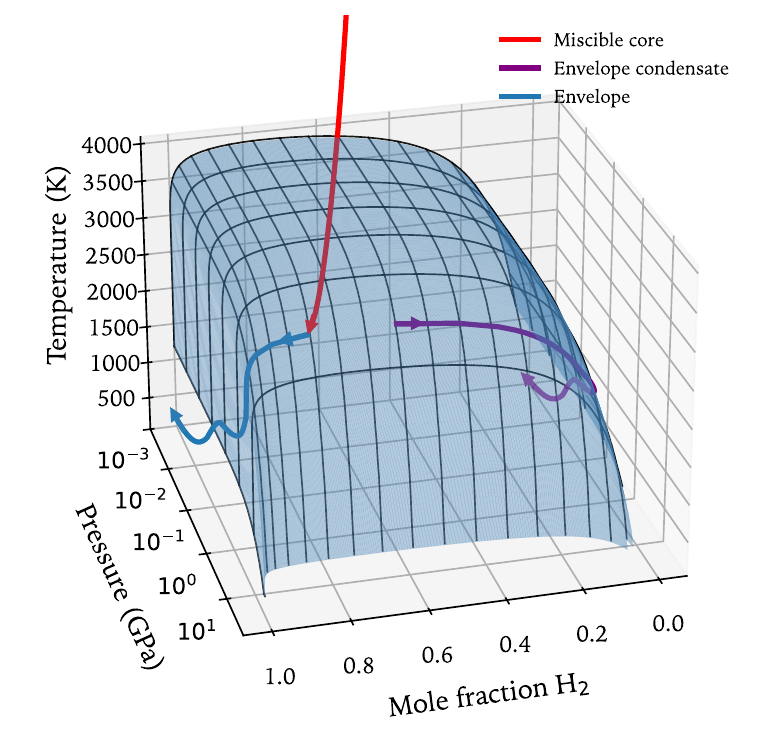}
  \caption{Plot of the binodal phase boundary in $T$, $P$, and mole fraction of H$_2$ space that separates the miscible interior and the outer envelope of Neptune as modeled here. The binodal phase boundary is shown as the blue dome structure. The curves terminating in arrow heads represent the miscible interior (red), H$_2$-rich gas (light blue), and liquid silicate rain (violet) with the arrow heads pointing in the direction of greater radial position. The binodal is calculated using the MgSiO$_3$-H$_2$ system \citep{gilmore_core-envelope_2025}, a reasonable approximation to the ternary MgSiO$_3$-Fe-H$_2$ system for this purpose \citep{Young2025_Differentiation}.}
  \label{fig:binodal_path}
\end{figure}

The influence of the binodal phase boundary on the structure and chemistry
of the model Neptune is illustrated in Figure~\ref{fig:binodal_path}.
The red curve in the figure is the adiabat within the miscible interior.
This path intersects the binodal phase-boundary surface in
$T$, $P$, and mole-fraction-of-H$_2$ ($x_{\rm H_2}$) space
(blue dome), at which point the miscible silicate--hydrogen--iron mixture
begins to exsolve into the two phases comprising the envelope: an
H$_2$-rich phase that becomes gas at low pressures, and liquid
condensates that we model as silicate rain that settles back to the base of the envelope. At first contact with the
binodal, the interior and incipient envelope are compositionally continuous and the first condensates are displaced to lower H$_2$ concentrations.
As $T$ decreases moving outward into the envelope, the silicate
liquids condense out, and in our model are removed from the envelope by settling. With further decreases in $T$ and $P$ higher in
the envelope, the relative mass of silicate liquid that rains out of the envelope increases, while the
condensate itself rapidly becomes depleted in H$_2$ (violet curve),
leaving the coexisting  H$_2$-rich fluid phase increasingly enriched in hydrogen
(light blue curve).

\subsection{Uranus}
The results for Uranus are broadly similar to those for Neptune but indicate a greater total hydrogen mass fraction
(Figure~\ref{fig:Neptune_and_Uranus}).  The input parameters are the total mass of
$14.536\,M_\oplus$ and the equilibrium temperature of 58~K.  

\begin{table}
\centering
\caption{Best-fit solution from the Uranus interior MCMC.
The reduced $\chi^2$ is computed with $n_{\rm obs} = 6$ constraints
and $n_{\rm free} = 3$ free parameters. A one-sided prior on the
melt-layer H$_2$ content (ceiling $12.0$~wt\%, $\sigma = 1.0$~wt\%)
is included but inactive at the best fit (predicted $10.41$~wt\%).}
\label{tab:uranus_bestfit}
\begin{tabular*}{\columnwidth}{@{\extracolsep{\fill}}lccc@{}}
\hline\hline
Quantity & Predicted & Target & $\chi^2$ \\
\hline
$P_{\rm binodal}$ (GPa)   & 7.04                 & ---                                  & --- \\
Bulk H$_2$ (wt\%)         & 14.36                & ---                                  & --- \\
$Ra/Ra_{\rm crit}$             & $5.674\times10^{9}$  & ---                             & --- \\
\hline
$J_2$ ($\times 10^{-6}$, static) & 3522.80       & $3522.76 \pm 13.470$               & 0.000 \\
$J_4$ ($\times 10^{-6}$, static) & $-30.880$     & $-32.062 \pm 3.000$                & 0.155 \\
$C/MR^2$                         & 0.2275        & $0.2250 \pm 0.00450$               & 0.309 \\
$R_{\rm eq}$ (km)                & 25561.7       & $25559.0 \pm 20.0$                 & 0.019 \\
$T_{\rm 1\,bar}$ (K)             & 75.0          & $76.0 \pm 5.0$                     & 0.040 \\
$L_{\rm int}$ (W)                & $6.65\times10^{14}$ & $6.30 \pm 1.5\times10^{14}$ & 0.054 \\
\hline
\multicolumn{3}{@{}l}{$\chi^2_{\rm total}$}                                              & 0.577 \\
\multicolumn{3}{@{}l}{$\chi^2_{\rm red}$ ($\nu = 3$)}                                    & 0.192 \\
\hline\hline
\end{tabular*}
\end{table}

The maximum
likelihood solution is obtained for a binodal pressure of $7.04$~GPa, a bulk planet H$_2$ fraction of $14.36\%$ by mass (as in the case for Neptune, this includes the nominal He
fraction by virtue of the adopted molecular weight for the non-metal fraction), and a $Ra/Ra_{\rm crit}$ value corresponding to a model intrinsic luminosity of $6.65\times 10^{14}$ W, indistinguishable from recent estimates (Table \ref{tab:uranus_bestfit}).  The reduced $\chi^2$ for the fit of $0.2$ is based on three degrees of freedom.

The Uranus uncertainties used for the MCMC searches follow a similar scheme as for Neptune, with values adjusted to reflect the different observational sources and model systematics.  For $J_4$, we adopt $\sigma_{J_4}$ equal to the magnitude of our wind-correction $\Delta J_4$ (Table~\ref{tab:uranus_constraints}), which is approximately seven times the Monte-Carlo uncertainty of \citet{French2024} but reflects the systematic uncertainty introduced by the wind correction itself. We similarly broaden $\sigma_{J_2}$ from the observational value of $4.12 \times 10^{-7}$ to the magnitude of our wind correction, $|\Delta J_2| = 1.35 \times 10^{-5}$.  The wind correction is more than thirty times the observational uncertainty in $J_2$, and even a modest fractional uncertainty in the wind-induced contribution \citep{French2024} dominates the formal observational error. This broadening of $\sigma_{J_2}$ is specific to Uranus but parallels the treatment of $J_4$ for both planets ($\sigma_{J_4} = |\Delta J_4|$).  For $R_{\rm eq}$, we adopt $\sigma_R = 20$~km, obtained by adding in quadrature the disparate  equatorial and polar 1-bar radius uncertainties reported by \citet{Lindal1987} ($\pm 4$~km and $\pm 20$~km respectively). For $T_{\rm 1\,bar}$, we adopt $\pm 5$~K rather than the Lindal observational $\pm 2$~K, matching the Neptune treatment.  For $C/MR^2$, we adopt a 2\% model uncertainty \citep{MovshovitzFortney2022, Siebenaler2026}, noting that $C/MR^2$ is not directly measured for Uranus and the adopted value of 0.2250 derives from \citet{Nettelmann2013}. Five of the six parameters fit by the MCMC match their target values within the original observational $1\sigma$ values, regardless of whether those values were increased for the search to include potential model systematic errors.  The exception is $J_4$, which deviates from the wind-corrected value by $2.5$ times the cited observational uncertainty (see Tables \ref{tab:uranus_bestfit} and \ref{tab:uranus_constraints}), although it is well within potential model uncertainties. This is not surprising given inherent uncertainties in modeling the dynamic effects on $J_{2n}$ values.  

The core density of Uranus in our models is less than that for Neptune by about $12\%$ due to the larger mass fraction of H$_2$ in the core (Table \ref{tab:nep_ura_compare}). Overall these results show that within this model structure, Uranus is ``puffier" than Neptune, commensurate with an overall greater mass fraction of hydrogen.  At the same time, the model Ledoux-stable boundary layer is thicker than that for Neptune, with a total boundary layer depth of $186$ km for Uranus compared with $129$ km for Neptune.  This is a physical explanation for the lower intrinsic luminosity for Uranus relative to Neptune provided by our model framework, but is not a prediction.  Rather, it is prescribed by our best-fit Rayleigh number ratio.   

\begin{figure*}[!t]
\centering
\includegraphics[width=\textwidth]{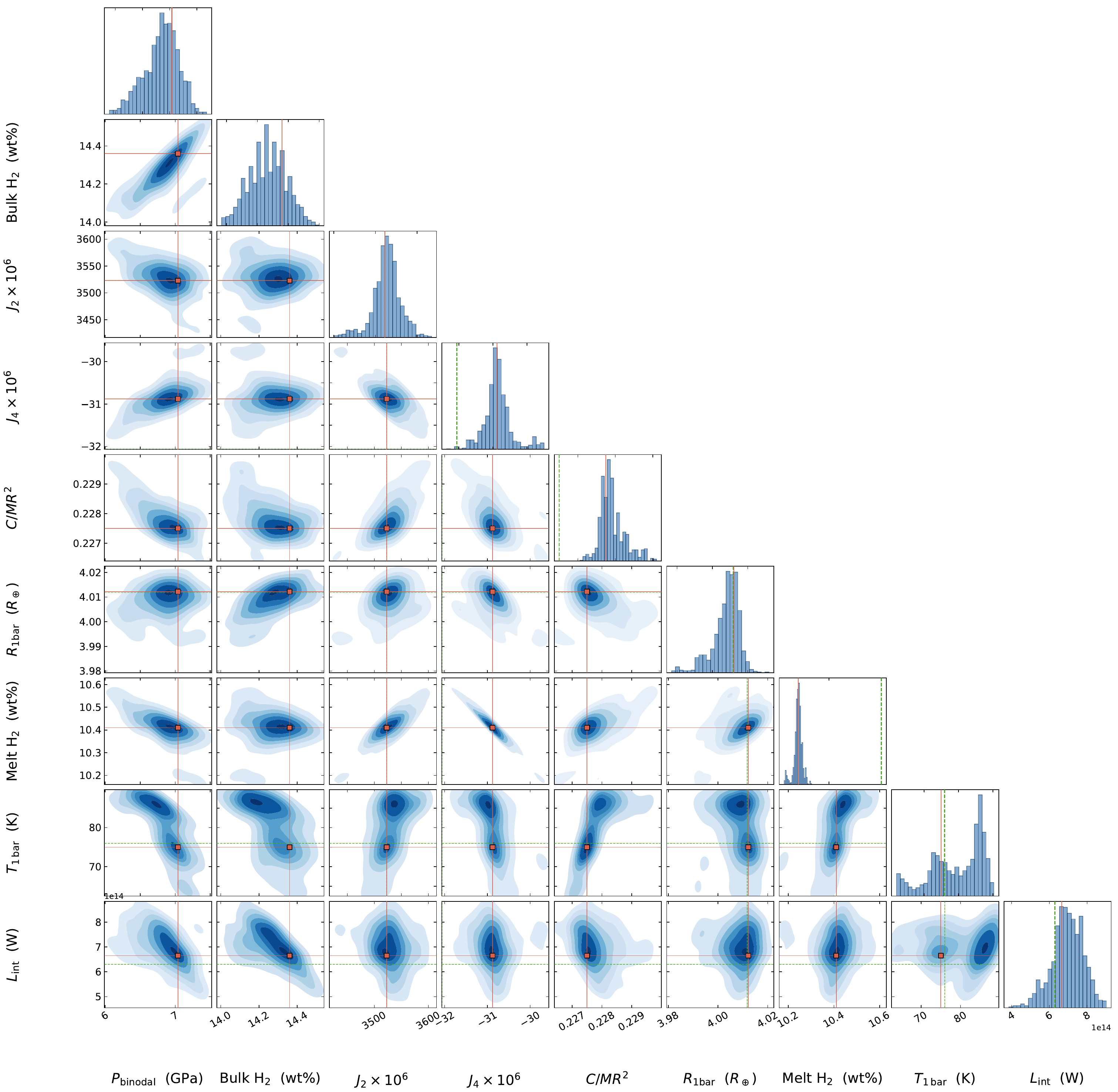}
\caption{Corner plot for MCMC search over 1,008 models for the best-fit for Uranus showing the marginalized posterior densities for the parameters of interest. The melt H$_2$ weight fraction that was not part of the fit is shown for reference. Green dashed lines are target values, and the red squares and red lines show the highest likelihood solution. }
\label{fig:Uranus_corner}
\end{figure*}

\section{Discussion}
\label{sec:discussion}
When assessing the quality of the fits obtained in these magma-ocean models, it is useful to distinguish between a constrained inverse structural problem, in which observed planetary properties are imposed as constraints on an assumed interior architecture, and a forward physical model, in which a specified structure and set of physical parameters generate observable quantities that are then compared with the data. In many classical ice-giant models, the inverse problem consists of prescribing observed bulk and atmospheric properties, such as $M_{\rm p}$, $R_{1,{\rm bar}}$, $T_{1,{\rm bar}}$, and $\omega$, and then solving for internal structural parameters that reproduce the measured gravity field. For example, \citet{Nettelmann2013} use this framework to constrain the metallicities of two envelope layers $Z_1$ and $Z_2$, the core mass $M_{\rm c}$, and the pressure $P_{1-2}$ at the transition between the two envelope layers for their model structure $\mathcal{M}$. Formally, this may be viewed as a constrained inverse problem in which one solves for $\Theta_{\rm model}$ such that
$\mathbf{d}_{\rm pred}(\Theta_{\rm model},\,\mathbf{x}_{\rm fixed}) = \mathbf{d}_{\rm obs}$, where
$\Theta_{\rm model} = \{Z_1, Z_2, M_{\rm c}, P_{1\text{-}2}\}$, the fixed inputs are $\mathbf{x}_{\rm fixed} = \{M_{\rm p}, R_{\rm 1\,bar}, T_{\rm 1\,bar}, \omega\}$, and the data to match are the gravity harmonics $\mathbf{d}_{\rm obs} = \{J_2, J_4\}$. The fixed inputs are in effect boundary conditions rather than data to fit, so by matching two gravity harmonics $\mathbf{d}_{\rm obs} = \{J_2, J_4\}$ while solving for the four structural features comprising $\Theta_{\rm model}$, the system is underdetermined by two degrees of freedom.

Because the physical chemistry of miscibility links structure and chemistry in our models, our approach is subtly different. In our calculation the observables are not imposed. Rather, for each trial set of fit parameters, the model returns a forward prediction for the full observable data vector, and the quality of the fit is determined by the likelihood $\mathcal{L}(\Theta_{\rm model})=p({\bf d}_{\rm obs}\mid \Theta_{\rm model},\mathcal{M})$ given the magma ocean model $\mathcal{M}$ with fit parameters $\Theta_{\rm model} = \{P_{\rm binodal}, xH_2, Ra/Ra_{\rm crit}\}$. Simultaneous agreement between the predicted and observed values, ${\bf d}_{\rm obs} = \{R_{\rm 1\, bar}, T_{1,{\rm bar}}, J_2, J_4, C/MR^2, L_{\rm int}\}$, is therefore not enforced a priori, but is instead a consequence of finding regions of parameter space with high likelihood.
This predictive approach has precedent in the ice-giant literature. \citet{BaileyStevenson2021} and \citet{CanoAmoros2024} let H$_2$--H$_2$O immiscibility physics determine the location of compositional gradients, rather than imposing them as free parameters, and \citet{Scheibe2021} computed the 1-bar radius as a forward output of thermal-evolution calculations, retaining models that did not match $R_{1\,{\rm bar}}$ to illustrate the physics. They invoked a thermal boundary layer similar in concept to that at the base of the H$_2$ envelope in this study.  Our approach is in the spirit of this same Bayesian model evaluation, in which the data are used to assess the likelihood of a specified physical hypothesis, rather than to  invert for the parameters of an assumed structure.

It is in this context that we can evaluate the fits of the models to both planets.
The model reproduces Uranus' six target observable values to well within their respective target uncertainties that include potential systematic errors, and five of the six are within purely observational uncertainties, the very tight nominal observational uncertainty for $J_4$ being the exception. The model also reproduces Neptune's observables to within their observational uncertainties, with a single exception: the predicted
1-bar temperature is high by approximately 12~K.

The $\sim 12$~K excess in $T_{1\,\rm bar}$ for Neptune is consistent
with potential systematic uncertainties in the grey thermal opacity $\kappa_{\rm th}$ used in our atmosphere model, exacerbated by $L_{\rm int}$. In the two-stream grey solution at large optical depth, we have
\begin{equation}
T_{1\,\rm bar}^4 \;\simeq\; T_{\rm eq}^4 +
  \frac{3\,L_{\rm int}\,\kappa_{\rm th}\,P_{1\,\rm bar}}
       {16\pi\,\sigma\,G M}.
\label{eq:T1bar_grey}
\end{equation}
\noindent While $T_{1\,\rm bar}$ scales weakly with opacity ($T_{1\,\rm bar}\propto\kappa_{\rm th}^{1/4}$ in the internal-heating term), the underlying $\kappa_{\rm th}$ uncertainties are large.  As an example, an unconstrained difference in the Rosseland-mean opacity of $2.5\times$, well within the spread of plausible grey averages for H$_2$--H$_2$ collision-induced absorption (CIA), CH$_4$ near-infrared bands, and aerosol contributions \citep{Valencia_2013}, produces a $\sim 15$~K shift in $T_{1\,\rm bar}$ that is comparable in magnitude to the residual observed for Neptune.  We interpret the $T_{1\,\rm bar}$ excess as the signature of inherent uncertainties in atmosphere-specific opacities rather than a clear structural defect in the interior model. 

That the miscible magma ocean model satisfies the joint constraints of bulk density, intrinsic luminosity, atmospheric thermal structure, and gravity harmonics for Uranus is notable in light of the long-recognized difficulty of constructing interior models that simultaneously reproduce these quantities \citep{HubbardMarley1989, HelledFortney2020}.  We emphasize that our model reproduces the observed low $L_{\rm int}$ by fitting the width of the boundary layer at the base of the atmosphere and so it is not a prediction of the model. The long-standing question of why Uranus is so much fainter than Neptune \citep{Kurosaki2017, Scheibe2021} remains an open question.  Our result does lend additional impetus to the concept of compositional gradients suppressing convection in the envelope \citep{VazanHelled2020} as the source of the low luminosity. 

Aside from simply providing another model for Uranus and Neptune that fits their observed characteristics, the magma ocean giant model has several ancillary features that recommend it.  One is that it highlights a potential connection between gas dwarf planets in general, from sub-Neptunes to Neptunes, with no requirement for a fundamental distinction in their formation mechanisms and resulting structures.  There is merit to deriving successful models that depend as much as possible on physical chemistry and material properties; it is preferable for the equations of state to do the work of matching the planet properties. Another feature to recommend the magma ocean giant model is that the basic features of the planets are reproduced using just three tunable parameters, pressure at the binodal, the bulk hydrogen fraction, and the $Ra/Ra_{\rm crit}$ ratio for the envelope that influences the intrinsic luminosity (i.e., the boundary layer).  The Bayesian information criterion (BIC) applied to Bayes factors is an example of a quantitative means of assessing the relative merits of complexity versus parsimony in model selection \citep{Aho2014}. While fewer parameters is not definitive evidence for veracity, it is a benefit when choosing among models based on material properties. Additional evidence for the presence of extant supercritical silicate-hydrogen magma oceans in the interiors of Uranus and Neptune and connections to extrasolar planets are discussed in what follows.  

\subsection{Atmosphere chemistry}
The atmospheric chemistries of both Uranus and Neptune are consistent
with predictions for magma ocean worlds. James Webb Space Telescope
(JWST) observations of the sub-Neptunes K2-18b and TOI-270d detect
abundant CH$_4$ and CO$_2$ but no NH$_3$ \citep{Madhusudhan2023_K2-18b,
Benneke2024, Holmberg2024_TOI270d}. NH$_3$ depletion is a natural
consequence of the high solubility of nitrogen in silicate melt under
the reducing conditions imposed by a thick H$_2$ envelope in contact
with a magma ocean \citep{Shorttle_2024, Werlen2026}, precisely the scenario
modeled here, and the atmospheric C/O ratio of such planets is set
by equilibrium with the underlying melt rather than inherited from
the protoplanetary disk, with CH$_4$ being the dominant carbon carrier in
the deep atmosphere \citep{werlen_atmospheric_2025}. The non-detection
of NH$_3$ in both Uranus and Neptune \citep{dePater1991} and their
observed CH$_4$ abundances are hence qualitatively consistent with magma
ocean interiors, lending independent support for the hypothesis that
the Solar System ice giants belong to the same class of objects as
the sub-Neptune gas dwarfs characterized by JWST. 

To test this quantitatively, we applied the
\texttt{ChemicalGlobalEquilibrium} code\footnote{\url{https://github.com/ExoInteriors/GlobalChemicalEquilibrium_Release}}
\citep{Grimm2026}, a global chemical equilibrium code originally
developed for super-Earths and sub-Neptunes
\citep{Schlichting_Young_2022}, to the conditions at the binodal
interface in our models. The code equilibrates a hydrogen-rich
atmosphere with coexisting silicate and metal phases. For the ice
giants, we exclude a separate metal phase, since metal is expected
to be miscible with the silicate-rich interior under Uranus- and
Neptune-like conditions \citep{Young2025_Differentiation}. This reduces the problem to a two-phase silicate--gas system. We base our chemical network on the one used
by \citet{Werlen2026}, which includes sulfur-, nitrogen-, and
oxygen-bearing species, but exclude the metal phase in our
calculations.

The mantle composition is dominated by MgSiO$_3$ (mole fraction
0.9211), with trace contributions from MgO, SiO$_2$, FeO, FeSiO$_3$,
Na$_2$O, and Na$_2$SiO$_3$. The gas phase is initialized with carbon,
sulfur, and nitrogen in solar proportions (CO $= 5.4\times10^{-4}$,
H$_2$S $= 2.64\times10^{-5}$, N$_2$ $= 6.76\times10^{-5}$). The
resulting equilibrium mixing ratios above the magma ocean are
dominated by H$_2$ ($x = 0.9782$) and H$_2$O ($x = 0.0613$), with
CH$_4$ ($x = 0.0149$) the principal carbon carrier, H$_2$S ($x =
5.08\times10^{-4}$) the principal sulfur carrier, and
nitrogen-bearing species essentially absent (NH$_3$: $x =
4.16\times10^{-9}$; N$_2$: $x = 4.16\times10^{-9}$).

These results reproduce the key observational signatures of the ice
giant atmospheres. CH$_4$ is the dominant carbon carrier with an
enrichment of $\sim$$28\times$ solar relative to hydrogen, broadly
consistent with the $40$--$50\times$ and $\sim$$80\times$ solar
enrichment observed in Uranus and Neptune respectively. Sulfur
appears primarily as H$_2$S at $20\times$ solar enrichment, matching
the $20$--$30\times$ solar values inferred for both planets
\citep{dePater1991}. Nitrogen-bearing species are virtually absent,
consistent with the non-detection of NH$_3$ in both planets. 
The model also produces water above the magma ocean, but it is expected to remain
unobservable.
Even under vigorous vertical mixing -- which can loft water-rich vapor 
upward from the deep atmosphere -- the steep decline of water's
saturation vapor pressure with temperature imposes a cold trap: rising
atmosphere saturates and condenses at the level (hundreds of bar) where the
temperature profile crosses the water condensation curve, limiting the
water partial pressure at its low saturation value regardless of the
mixing strength embodied by eddy diffusion $K_{zz}$, for example. CH$_4$, by contrast, has a high enough
saturation vapor pressure at relevant low temperatures to persist as a
detectable vapor in the observable 1 to 10~bar region 
\citep{Fray2009}. The equilibrium chemistry
of a magma ocean interface therefore reproduces the observed
pattern of CH$_4$ and H$_2$S enrichment together with nitrogen depletion,
without requiring fine-tuning of atmospheric or disk conditions.
Of course non-detection in the upper atmospheres
is not proof of extant magma oceans nor is it proof of depletion of the atmospheres as a whole \citep{Li_2017, Guillot_2019}, but it is
consistent with expectations where magma oceans are present.

\subsection{Hydrogen accretion and links to sub-Neptunes}
Later stages of planet formation are likely dominated by the merging
of planetary embryos, each with their own H$_2$-rich primary
atmospheres. The observed spread in inferred H$_2$ mass fractions
among sub-Neptunes spans roughly an order of magnitude, from $\sim
1\%$ to $\sim 10\%$ by mass \citep{Bean2021}, providing some clue as
to the connection between Neptune and Uranus and the broader
sub-Neptune population.

To explore this connection, we performed a simple stochastic numerical
accretion experiment in which planets of five target masses ($2$, $5$,
$10$, $15$, and $20\,M_\oplus$) are assembled by sequential addition
of embryos drawn at random from a log-uniform mass distribution
(minimum mass $0.1\,M_\oplus$, maximum mass half the target mass),
reflecting the embryo-dominated late stages of accretion. The
characteristic disk temperature at the planet's location is assigned
to reflect formation distance, ranging from 300~K for $2\,M_\oplus$
to 50~K for the most massive planets motivated by the architecture of the Solar System. The precise radial temperature profile is not essential to the salient features of the calculations presented here. During each accretion event two processes occur. First, the impact partially strips the pre-existing H$_2$ envelope, so that only a fraction $x_{r_i}$ of the envelope mass survives. Second, the
accreted embryo contributes additional H$_2$ according to the
cooling-limited scaling of \citet{ginzburg2016a},
\begin{equation}
x_{\rm H_2}(m) =
0.02
\left(\frac{m}{M_\oplus}\right)^{0.8}
\left(\frac{T}{1000\,{\rm K}}\right)^{-1/4}
\left(\frac{t_{\rm disk}}{1\,{\rm Myr}}\right)^{1/2},
\end{equation}
where $T$ is the characteristic disk temperature and $t_{\rm disk}$
is the gas disk lifetime. The envelope mass therefore evolves as
\begin{equation}
M_{\rm env}^{(k+1)} = x_{r_k}\,M_{\rm env}^{(k)} + \Delta M_k,
\label{eqn:Mk}
\end{equation}
where $\Delta M_k = f(m_k)\,m_k$ is the H$_2$ increment contributed
by the $k$-th embryo of mass $m_k$. Iterating through $N$ accretion
events gives
\begin{equation}
M_{\rm env}^{(N)} =
\left(\prod_{j=0}^{N-1} x_{r_j}\right) M_{\rm env}^{(0)}
+
\sum_{k=0}^{N-1}
\left[
\Delta M_k
\prod_{j=k+1}^{N-1} x_{r_j}
\right],
\label{eqn:MN_iterated}
\end{equation}
so that the final envelope is the sum of all accreted increments,
each weighted by the retention factors of all subsequent impacts.
The final H$_2$ mass fraction is then
\begin{equation}
x_{\rm H_2} = \frac{M_{\rm env}}{M_{\rm core} + M_{\rm env}}.
\end{equation}
Because the retention factor $x_{r_k}$ enters multiplicatively at
each step, the central limit theorem implies that $\ln(M_{\rm env})$
behaves approximately as a sum of random contributions, and the
resulting H$_2$ mass fraction distribution is approximately
log-normal. The width of the distribution at each target mass
reflects the stochastic variability in both the sequence of embryo
masses and the sequence of retention factors.

\begin{figure}[t]
  \centering
  \includegraphics[width=0.9\columnwidth]{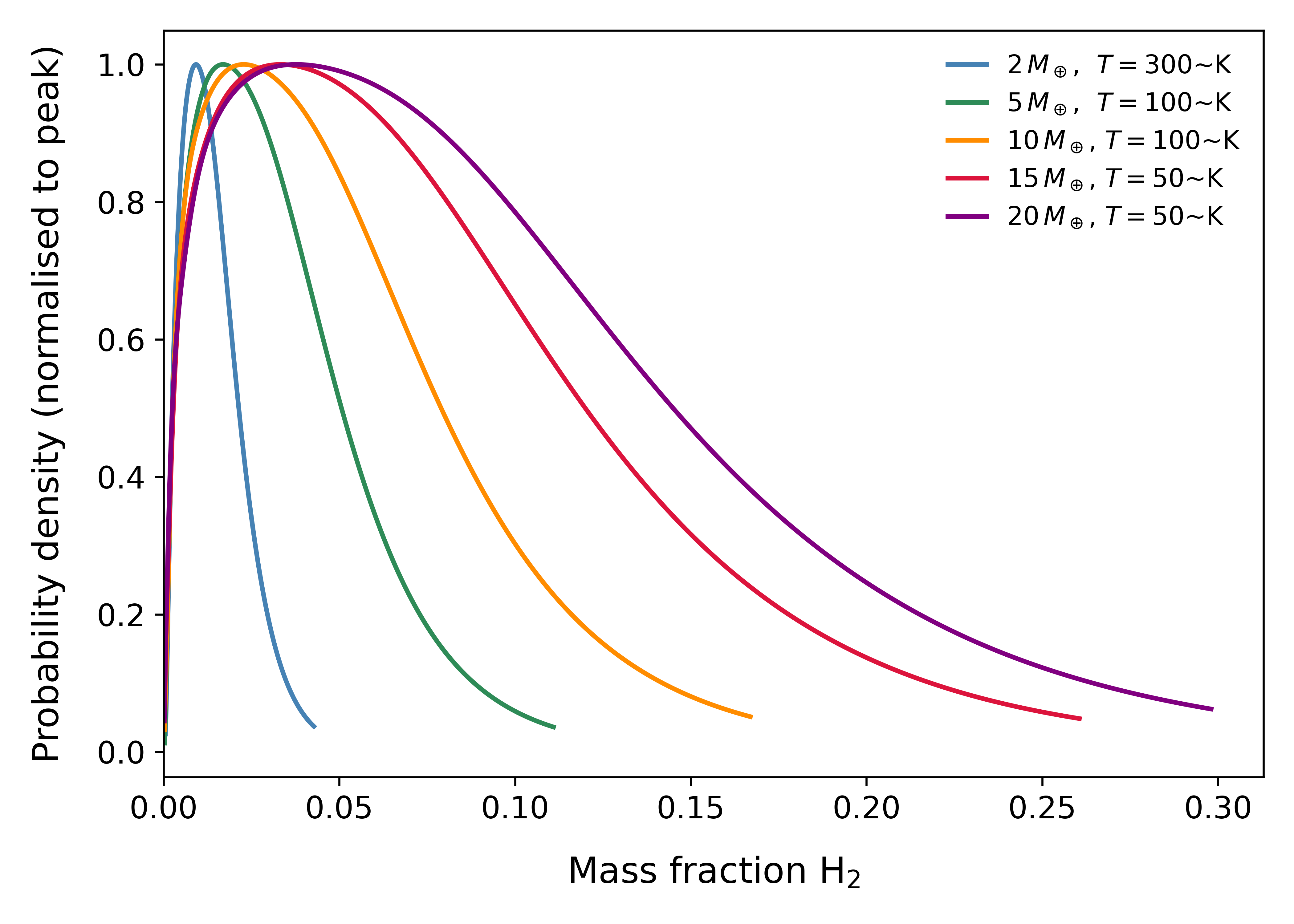}
  \caption{Results of a toy accretion model illustrating the expected relationship between planet mass and hydrogen fraction.}
  \label{fig:H2dist}
\end{figure}

The results, shown in Figure~\ref{fig:H2dist}, demonstrate that the
median H$_2$ fraction increases systematically with planet mass, from
$\sim 1\%$ for $2\,M_\oplus$ to $\sim 9\%$ for $20\,M_\oplus$. Two
effects drive this trend. First, and most importantly, more massive planets accrete larger
individual embryos, and the scaling $x_{\rm H_2} \propto m^{0.8}$
means each event contributes disproportionately more H$_2$ per unit
solid mass as embryo masses grow \citep{ginzburg2016a}. Second, more
massive planets form at cooler disk temperatures in this model,
further boosting the accreted envelope fraction via the $T^{-1/4}$
dependence. The $\sim 13\%$ H$_2$ mass fractions inferred for Uranus
and Neptune from our interior models are fully consistent with this
picture, sitting at the high-mass end of the predicted distribution.
The magma ocean giant model is thus both parsimonious --- reproducing
the structure, chemistry, and thermal states of both planets with
just three free parameters --- and physically motivated, connecting
directly to the broader population of gas dwarf planets and their
formation.

\subsection{Implications for magnetic fields}
 
Measurements of Uranus' and Neptune's magnetic fields also suffer from the limitations of only single flybys by Voyager II. However, despite only having information on the large-scale structure of their fields, it is evident that they differ dramatically from those of the two gas giants, and Earth, as they are not dominated by dipoles \citep{Ness1986, Ness1989}. Dynamo studies suggest that the generation of so-called ``multipolar" magnetic fields does not require a specific dynamo region geometry, or a more slowly rotating system \citep[e.g.][]{Gastine_2012b, Yadav_2013, Soderlund_2025}.

DFT-MD simulations have shown that the magnetic Reynolds numbers for molten silicates at relevant conditions are sufficient to produce magnetic fields \citep{soubiran_2018}. Silicate melts enriched in iron exhibit a further increase in magnetic Reynolds number \citep{Dragulet_2025}, and we anticipate that the presence of hydrogen would amplify this effect. Thus, there is no reason to doubt that multipolar magnetic fields may be generated in the models presented here. 

\subsection{Evolution}

We calculated the time evolution of Neptune based on our model.  The
quasi-static evolution curves are obtained by specifying the planet
mass $M_p$, bulk hydrogen mass fraction, equilibrium temperature
$T_{\rm eq}$, and an initial binodal pressure that results in the model
binodal pressure today. The binodal pressure is 
uniquely correlated with total energy and thermal state.  For each step we calculated the total
energy of the planet $E_p$ and the associated intrinsic luminosity
$L_{\rm int}$.  The timestep associated with a step from $P_i$ to
$P_{i+1}$ was obtained from energy balance,
\begin{equation}
dt = \frac{\left|E_p(P_{r_{i+1}})-E_p(P_{r_i})\right|}{L_{\rm int}},
\end{equation}
where $L_{\rm int}$ is output from the prior time step.  Total energy
$E_p$ is calculated from
\begin{equation}
E_{p} = \int_{0}^{R_{p}}
\left(-\frac{G M(r)}{r} + c\,T(r)\right)\,\rho(r)\,4\pi r^2\,dr,
\label{eqn:E_total}
\end{equation}
where $c$ is the specific heat capacity for the melt or envelope using
the equations of state and species described above.  Pressure steps
are sufficiently small to ensure that $dt$ is approximately four to
five orders of magnitude smaller than the cooling time constant
$\tau_{\rm cool} = |E_p|/L_{\rm int}$.  

\begin{figure*}[!t]
\centering
\includegraphics[width=0.9\textwidth]{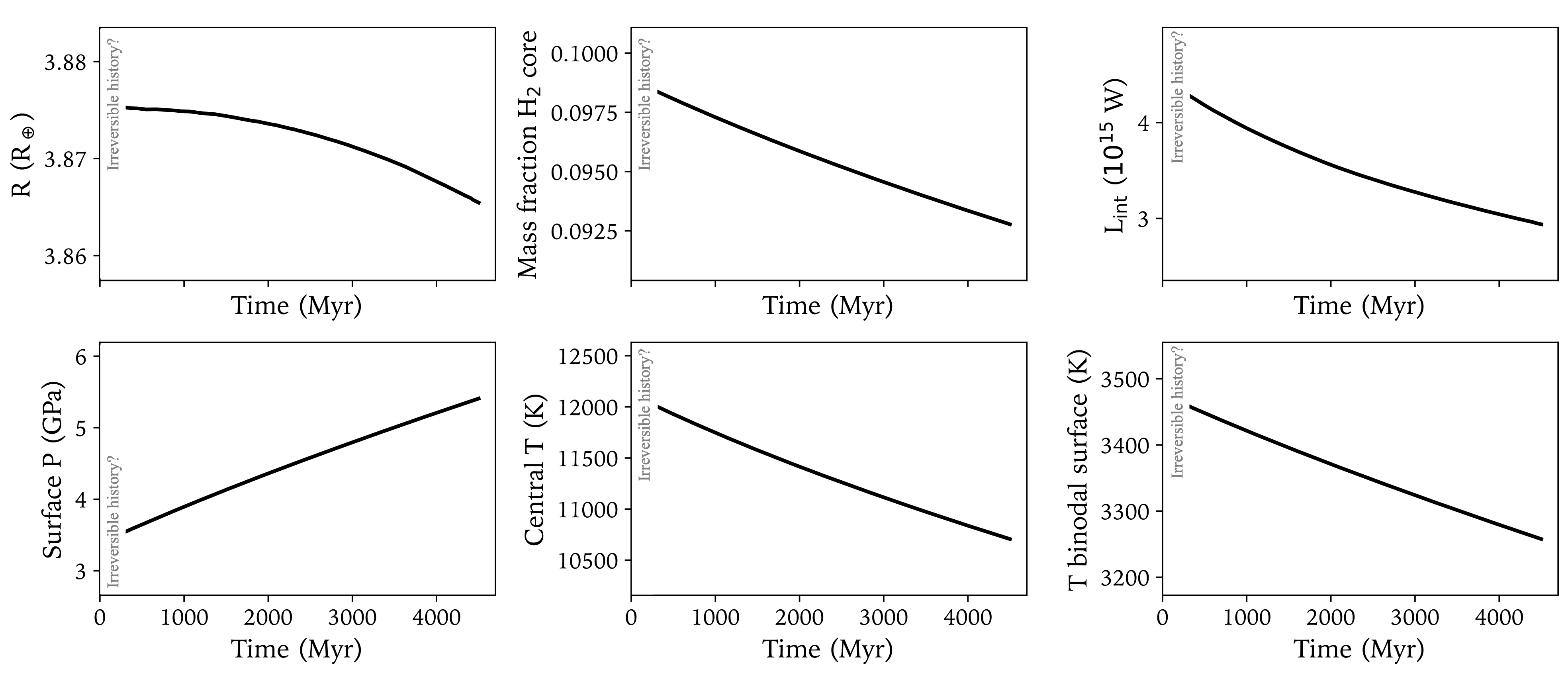}
\caption{Calculated evolution of Neptune over the last $4.5$~Gyr. Irreversible processes that predate the present-day structure of the planet are not resolved with this quasi-statical calculation. 
}
\label{fig:evolution}
\end{figure*}

Results indicate that the changes in radius, temperatures, and partitioning of hydrogen have been relatively modest over the last $4.5$ Gyr (Figure~\ref{fig:evolution}). Comparing the temporal changes in radius with the mass fraction of H$_2$ in the core illustrates that the exsolution of hydrogen plays a role in limiting the decrease in radius during cooling.  Unknown, highly irreversible events that may have occurred very early in the formation of the planet are not included here by necessity. 

\subsection{Limitations}
The specifics of the magma ocean giant models for Uranus and Neptune are not unique. Our numerical experiments show that reasonable changes to the supercritical melt EoS parameters result in similarly successful fits to the planet observables but with similar, although still distinct, $P_{\rm binodal}$ and $x_{\rm H_2}$ values. These models with different binodal pressures and bulk hydrogen fractions have essentially the same pressure-density curves as those obtained here. The precise parameters that determine these curves are affected by the supercritical melt EoS (see Appendix D).    The uniqueness of the solutions will be improved as the equation of state for miscible MgSiO$_3$, H$_2$, and Fe becomes better constrained (work in progress).  Until then, we can conclude with confidence that the observable characteristics of these planets can be fit with extant, supercritical hydrogen-rich magma oceans comprising the inner 3/4 of their radii, a boundary layer stable against convection due to molecular weight gradients above the magma oceans, and an overlying hydrogen-rich envelope.  The precise concentrations of hydrogen, and precise pressures and temperatures at the interfaces between the magma oceans and the envelopes will require refinement of the melt phase EoS. For now, we can say that there is a systematic absolute uncertainty of order $2\%$ (e.g., mass fraction of H$_2$ for the core of Neptune is $9.5\% \pm 2$) in the mass fraction of total hydrogen for both planets in our models based on uncertainties in the EoS for the supercritical melt phase (Appendix D).  

\section{Conclusions}
\label{sec:conclusions}
Modern cosmochemical and astrophysical constraints on the relative abundances of silicates and ices suggest that the progenitors of the ice giant planets in the Solar System may have been "rockier" than previously asserted. At the same time, our understanding of the physical chemistry of mixtures of rocky material and hydrogen at high temperatures and pressures has evolved.  Against this backdrop, we have the opportunity to use the ice giants as a test of models put forward previously for sub-Neptune planets.  We have asked and answered the question of whether the observable constraints for Neptune and Uranus can be explained by interiors composed of extant supercritical mixtures of silicate and hydrogen, overlain by H$_2$-rich envelopes.  This same basic structure has been proposed for sub-Neptunes for the same reasons; temperatures and pressures are sufficiently high in these bodies that miscibility between rock and hydrogen should obtain. 

The apparent disparity in the orbital distances from their host stars does not necessarily translate to fundamentally distinct interior structures for the Solar System ice giants and sub-Neptunes.  Accreted water ice is reworked by magma ocean-hydrogen envelope reactions, partly, if not entirely, erasing the signature of those ices \citep{werlen_sub-neptunes_2025, Werlen2026}. Based on the chemical model presented here and elsewhere, the primary manifestation of the progenitor ices will be to provide an additional source of hydrogen indistinguishable from that provided by H$_2$ and enhance the overall oxidation states of these planets, rather than altering their basic structure. 

While this is just one of a number of models that successfully reproduce the observed features of Neptune and Uranus, this model has several aspects to recommend it.  One is the connection with other gas dwarf planets; it is not clear that ice giants and sub-Neptunes should be fundamentally different simply because of their distances from their host star. Related to this is the fact that the most basic chemical features of the ice giants resemble those of gaseous sub-Neptunes, perhaps indicating similar boundary conditions for the chemistry of the atmospheres imposed by the magma oceans.  Another potential favorable feature of the magma ocean giant proposition is that the use of the phase equilibria involving silicates and hydrogen, arguably the dominant chemical constituents of the planets, limits the number of tunable parameters to reproduce their observed densities, external gravity, and calculated moments of inertia to just two. Simpler models are in many ways easier to test, and are less susceptible to overfitting.  However, more work is required to improve relevant equations of state and address the initial conditions for gas dwarf planet formation.

\section*{Appendix A: Planet Structure} \label{sec:Appendix_A}

Many of the methods for calculating the planet models used here have
been described previously \citep{Young2025_Differentiation,
RogersYoungSchlichting2025_MNRAS, young_influences_2026}.  However, there are some
modifications to the prior published work, and so the models are
explained here.  To derive the core, we solve the system of equations
\citep{Seager2007}:
\begin{equation}
\frac{dm}{dr}=4 \pi r^2 \rho,
\label{eq:dmdr}
\end{equation}
\begin{equation}
\frac{dP}{dr}=-\frac{Gm \rho}{r^2},
\label{eq:dPdr}
\end{equation}
and
\begin{equation}
\left(\frac{dT}{dr}\right)_S
= -\,\frac{\gamma(\eta)\,T}{K_S(\eta)}\,\rho\,g,
\label{eq:dTdr}
\end{equation}
where $m$ is the mass contained within radius $r$, $\rho$ is mass
density, $\gamma(\eta)$ is the Gr\"uneisen parameter that varies with
$P$ and $T$ through the compressibility factor $\eta = \rho/\rho_0$,
and $K_S(\eta)$ is the isentropic bulk modulus.  We assume the core is
fully convective, with a limiting isentropic $dT/dP$ gradient.  In
practice, we evaluate the bulk modulus using
$K_S(\eta) \equiv \eta\,d\!\left(P\!\left(\eta,T_S\right)\right)/d\eta$
where $P(\eta,T_S)$ is the total isentropic pressure along the adiabat
with temperature $T_S$.  Numerically integrating
Equations~(\ref{eq:dmdr})--(\ref{eq:dTdr}) through one or more layers
(with or without a metal core) yields a density and temperature profile
for the planet interior.

For the structure of the envelope we integrate numerically the
hydrostatic and mass conservation equations as in the interior
(Equations~\ref{eq:dmdr} and \ref{eq:dPdr}) using the tabulated function
from \citet{Chabrier2019} for the EoS.  The temperature structure of
the envelope is obtained by numerically integrating the coupled
hydrostatic, mass conservation, and energy transport equations outward
from the magma ocean surface, followed by an iterative inward
integration from the optically thin outer atmosphere.  Throughout most
of the envelope, energy transport is convective, and the temperature
gradient is therefore taken to be the convective (pseudoadiabatic)
gradient,
\begin{equation}
\frac{dT}{dr} = -\,\nabla T_{\rm conv},
\end{equation}
where $\nabla T_{\rm conv}$ is evaluated using a moist pseudoadiabat
appropriate for an H$_2$--silicate vapor mixture \citep{Graham_2021},
scaled to the H$_2$ equation of state \citep{Chabrier2019}.  Above the
convective layer where the gas is still collisional, radiative diffusion
applies,
\begin{equation}
\nabla T_{\rm rad}
=
-\frac{3 \kappa \rho L_{\rm int}}{64 \pi \sigma T^3 r^2},
\label{eqn:nablarad}
\end{equation}
where $\kappa$ is the local opacity, $\rho$ the density, and
$L_{\rm int}$ the intrinsic luminosity carried upward from the interior.
The envelope transitions smoothly into a radiative regime by relaxing
the temperature gradient with Eddington's two-stream solution for a
plane-parallel grey atmosphere \citep{Guillot2010},
\begin{equation}
\begin{aligned}
T^{4}(\tau)
&=
\frac{3}{4}\,T_{\rm int}^{4}
\left(\tau+\frac{2}{3}\right) \\
&+\frac{3}{4}\,T_{\rm eq}^{4}
\left[
\frac{2}{3}
+\frac{2}{3\gamma_{\rm vis}}
+
\left(
\frac{\gamma_{\rm vis}}{3}
-\frac{2}{3\gamma_{\rm vis}}
\right)
e^{-\gamma_{\rm vis}\tau}
\right],
\end{aligned}
\label{eqn:Eddington}
\end{equation}
where $\tau$ is the optical depth measured downward from the top of the
atmosphere, $T_{\rm eq}$ is the equilibrium temperature set by stellar
irradiation, $T_{\rm int}$ is the intrinsic temperature associated with
the interior luminosity, and $\gamma_{\rm vis}$ in this context is the
visible-to-thermal opacity ratio (we used 1 for these calculations, but
the results do not depend critically on this value).  We impose
conservation of intrinsic luminosity, rather than a constant intrinsic
flux as assumed in the plane-parallel equations, by accounting for the
area dependence of the intrinsic flux
$f_{\rm int}(r) = \sigma T_{\rm int}^4(r)$ such that
$T_{\rm int}(r) = T_{\rm int, s} (r_s/r)^{1/2}$, where subscript $s$
signifies evaluation just above the supercritical magma ocean.  At
large infrared optical depth ($\tau \gg 1$), Equation~\ref{eqn:Eddington}
approaches the radiative--diffusion limit, yielding a radiative
temperature gradient consistent with Equation~\ref{eqn:nablarad} and
thus ensuring consistency with deep-envelope energy transport.  This
approach affords a smooth and numerically stable transition from the
deep convective adiabat to the irradiated upper atmosphere.

\section*{Appendix B: Luminosity} \label{sec:Appendix_B}
Specifying an initial entropy and intrinsic luminosity, $L_{\rm int}$,
based on inferences about retention of heat during accretion does not
guarantee self-consistency; a specified $L_{\rm int}$ may exceed, or
fall short of, the flux the atmosphere can transmit given its opacity
structure and boundary conditions, leading to either interior heating
or unphysical cooling timescales.  A closure for this problem, and thus
self-consistent values for $L_{\rm int}$, can be obtained from
consideration of boundary layer effects.

We include the expected formation of a boundary layer at the base of
the hydrogen-rich envelope between the supercritical melt and the
overlying convective region in the envelope \citep{Ahlers_2009},
augmented by the hindrance of convection due to the mass loading of
heavy elements at relatively high temperatures \citep{Misener2023}.
The lower boundary layer restricts the intrinsic upward luminosity that
comprises heat transfer from the core to the atmosphere, forming a
thermal bottleneck.

The minimum thickness of the basal boundary layer separating the
interior (core) and the overlying envelope, $\delta$, comes from
consideration of the boundary for Rayleigh--B\'enard convection in the
envelope.  The width of this boundary is controlled by the vigor and
scale of the convection and is given by scalings similar to
\begin{equation}
\delta \sim n h\,(Ra/Ra_{\rm crit})^{-1/3},
\label{eqn:delta}
\end{equation}
where $h$ is a characteristic convecting length scale, taken to be some
multiple $n$ of the pressure scale height evaluated at the base of the
atmosphere, $Ra$ is the Rayleigh number, and $Ra_{\rm crit}$ is the
critical Rayleigh number for the convecting H$_2$-rich envelope
\citep{Long_20202, Grossmann_2000}.  The Rayleigh number is computed as
\begin{equation}
Ra = \frac{g\,\alpha\,\Delta T\,h^3}{\nu\,\kappa},
\label{eq:ra}
\end{equation}
where $h$ is here taken to be $5H$ for scale height $H$, $\Delta T$ is
the temperature contrast across the layer, $\nu$ is the kinematic
viscosity, $\kappa$ is the thermal diffusivity, and $\alpha$ is the
thermal expansivity.  Use of the adiabat for $\Delta T$, rather than
superadiabatic temperatures, likely overestimates $Ra$, resulting in
underestimates for the widths of the boundary layers (see below).
Kinematic viscosities for H$_2$ are estimated here using a Sutherland-law
fit for dynamic viscosities $\eta$:
\begin{equation}
\eta(T) = \eta_0 \left(\frac{T}{T_0}\right)^{3/2}
          \frac{T_0 + S_{T}}{T + S_{T}},
\label{eqn:dynamic_viscosity}
\end{equation}
with reference viscosity $\eta_0$ at reference temperature $T_0$ and
Sutherland temperature $S_T$ for H$_2$ \citep{Braun2018}.  The
kinematic viscosities are then
\begin{equation}
\nu = \frac{\eta}{\rho}.
\label{eqn:kinematic_viscosity}
\end{equation}
Thermal expansivities required to evaluate $Ra$ are obtained using
\begin{equation}
\alpha = -\frac{1}{T}\left(\frac{\partial\log\rho}{\partial\log T}\right)_{\!P}
\label{eqn:alpha}
\end{equation}
based on the EoS tables for hydrogen \citep{Chabrier2019}.  We adopt a
thermal diffusivity $\kappa$ of $1.0\times10^{-5}$~m$^2$~s$^{-1}$ for
the purpose of evaluating approximate values for the Rayleigh number.
For nominal parameters for Neptune, e.g., $\Delta T = 3000$~K, surface
pressure $P_s = 6$~GPa, planet mass $M_p = 17.1\,M_\oplus$, and density
$\rho = 230$~kg~m$^{-3}$, one obtains $Ra \simeq 10^{32}$.

The absolute value of $Ra_{\rm crit}$ is subject to large systematic
uncertainty.  Rotating convection theory predicts
$Ra_{\rm crit} \approx 8.7\,\mathrm{Ek}^{-4/3}$ as
$\mathrm{Ek}\to 0$ \citep{Chandrasekhar1961}, where the Ekman number is
$\mathrm{Ek}=\nu/(2\Omega H^2)$ and $\Omega$ is the angular rotation
rate.  At the base of the planetary envelope
$\mathrm{Ek} \sim 10^{-18}$, so that extrapolating the Chandrasekhar
scaling involves seven orders of magnitude beyond the regime where it
is known to hold.  The actual onset and vigor of convection in a
compressible, compositionally stratified planetary envelope is an open
theoretical problem.  The values of $Ra/Ra_{\rm crit}$ derived in our
models should therefore be understood as effective parameters that
reproduce the observed heat flows, rather than as first-principles
predictions from rotating convection theory.  For Neptune's rotation
rate we obtain a value for $Ra_{\rm crit}$ of $10^{22}$, and thus
$Ra/Ra_{\rm crit} \sim 10^{9}$.  However, reproducing Neptune's
observed internal luminosity $L_{\rm int} \sim 3\times10^{15}$~W
requires $Ra/Ra_{\rm crit} \simeq 10^{12}$, yielding
$\delta \approx 100$~m.  The analogous fit for Uranus, whose internal
heat flux is several times smaller, requires $Ra/Ra_{\rm crit}
\simeq 10^{10}$, yielding $\delta \approx 700$~m.  The physically
meaningful and robust result is the \emph{ratio}
$(Ra/Ra_{\rm crit})_{\rm Neptune} / (Ra/Ra_{\rm crit})_{\rm Uranus} \sim 100$.  The ratio is independent of the absolute calibration of the Chandrasekhar scaling.  The factor of $\sim 100$ difference in criticality for convection between the two planets is physically
plausible if their deep envelopes differ in viscosity, detailed
composition, or superadiabatic temperature excess, all of which would
affect the boundary layer thickness.

The model intrinsic luminosity $L_{\rm int}$ is determined from the
blackbody emission $L_{\rm bb} = 4\pi r_{\rm s}^2 \sigma T_{\rm s}^4$
using the ability of the magma ocean to radiate through the boundary
layer immediately above its surface, which for a grey atmosphere is
\begin{equation}
L_{\mathrm{int}} = L_{\rm bb}
\left(\frac{4}{3(\tau_{\rm BL}+1)}\right).
\label{eqn:Lint}
\end{equation}
The intrinsic temperature at the surface satisfies
\begin{equation}
L_{\rm int} = 4 \pi r_{\rm s}^2 \sigma T_{\rm int, s}^4.
\label{eqn:Tint}
\end{equation}
This formulation allows for self-consistency between the surface of the
magma ocean and the upper atmosphere, accommodating the link between
intrinsic luminosity in the regime of vigorous convection.  Rather than
resolving the temperature gradient in this often very narrow boundary,
its influence as a governor on radiative flux is captured through
Equation~\ref{eqn:Lint}.

The basal radiative layer is further stabilized and enhanced in
elevation by the molecular weight gradients that attend the interaction
between the supercritical magma ocean and the overlying envelope.  As
a result, the boundary between the surface of the magma ocean and the
convective layer in the envelope can be both pronounced and explicitly
resolvable in our models.  Rather than allowing arbitrarily large
intrinsic luminosities that would dynamically destroy stratification,
we adopt a quasi-static closure in which a Ledoux-stable layer, when
permitted thermodynamically, forms, persists, and therefore acts as a
finite thermal resistance that limits the transmitted flux.

Instability against convection in the presence of composition gradients
is governed by the Ledoux criterion:
\begin{equation}
\nabla_{\rm rad} > \nabla_{\rm ad} + \frac{\phi}{\delta}\,\nabla_\mu,
\label{eqn:Ledoux_criterion}
\end{equation}
where
\begin{align}
\nabla_\mu &\equiv \frac{d\ln \mu}{d\ln P}, \\[6pt]
\delta &\equiv -\left(\frac{\partial \ln \rho}{\partial \ln T}\right)_{P,\mu}, \\[6pt]
\intertext{and}
\phi &\equiv \left(\frac{\partial \ln \rho}{\partial \ln \mu}\right)_{P,T}.
\label{eqn:Ledoux}
\end{align}
Here $\mu$ is the mean molecular weight.  A positive molecular-weight
gradient ($\nabla_\mu > 0$) increases the critical temperature gradient
required for convective instability by the factor
$(\phi/\delta)\nabla_\mu$.  The ratio $(\phi/\delta)$ is unity for an
ideal gas and remains of order unity for reasonable values of the
compressibility factor $Z$, so in practice we simplify the Ledoux
criterion (Equation~\ref{eqn:Ledoux_criterion}) to
$\nabla_{\rm rad} > \nabla_{\rm ad} + \nabla_\mu$.  Where the
Ledoux-stable region spans a finite radial extent, it further limits
$L_{\rm int}$ for the overlying atmosphere layers.  The intrinsic
luminosity is limited by the requirement that the layer remain
marginally Ledoux-stable.  In this radiative layer, the upward energy
flux is given by the diffusion approximation, written in terms of
dimensionless gradient $\nabla$ as
\begin{equation}
F_{\rm rad}
=
\frac{4 \sigma T^{4}}{3 \kappa \rho H}\,\nabla,
\end{equation}
where $H=P/(\rho g)$ is the pressure scale height and
$\nabla \equiv d\ln T/d\ln P$.  The maximum radiative flux consistent
with stability against convection is obtained by evaluating this
expression at the largest temperature gradient permitted by the Ledoux
criterion:
\begin{equation}
\nabla_{\max} = \nabla_{\rm ad} + \nabla_{\mu}.
\end{equation}
In the radiative layer, we therefore have
\begin{equation}
\nabla_{\rm rad}= \nabla_{\rm max} =
\frac{3 \kappa P L}{64 \pi \sigma G M T^{4}},
\label{eqn:nabla_rad}
\end{equation}
leading to a definition for the maximum luminosity that can be
transported without violating Ledoux stability:
\begin{equation}
L_{\mathrm{Ledoux}}
=
\left(\frac{64 \pi \sigma G M T^{4}}{3 \kappa P}\right)
\nabla_{\rm max},
\label{eqn:L_ledoux}
\end{equation}
evaluated at a representative location near the top of the
Ledoux-stable region.  The intrinsic luminosity is then capped such
that $L_{\rm int} = \min(L_{\rm int}',\,L_{\mathrm{Ledoux}})$, where
$L_{\rm int}'$ is the value in the absence of the stable boundary.
This upper limit is applied iteratively during the solution procedure
until a self-consistent thermal structure satisfying Ledoux stability
is obtained.  Any larger luminosity would require a radiative
temperature gradient steeper than $\nabla_{\rm max}$, causing
$\nabla_{\rm rad}$ to exceed the Ledoux stability limit.  In that case
the stabilizing molecular-weight gradient would no longer suppress
convection, and the layer would become convectively unstable.

Expressing the radiative flux as a finite difference across the
Ledoux-stable region, we have
\begin{equation}
L_{\rm Ledoux}
\simeq
4\pi r_{\nabla_{\mu,\rm max}}^{2}\,
\frac{4\sigma}{3}\,
\frac{T_{\rm lower}^{4}-T_{\rm upper}^{4}}{\Delta\tau_{\rm Ledoux}},
\label{eqn:L_ledoux_tau}
\end{equation}
which makes explicit that the transmissible luminosity is controlled by
the integrated optical-depth thickness across the inhibited region
$\Delta\tau_{\rm Ledoux}$.  Here $T_{\rm lower}$ and $T_{\rm upper}$
are the temperatures at the lower and upper boundaries of the layer,
and $r_{\nabla_{\mu, \rm max}}$ is a representative radius within the
region where $\nabla \mu$ is maximized.  Equation~(\ref{eqn:L_ledoux_tau})
therefore represents the integrated radiative throughput of the
Ledoux-stable layer and provides a global maximum for the intrinsic
luminosity.  It also avoids numerical oscillations that can arise from
high-frequency layer-by-layer adjustments of the temperature gradient.

Molecular weight gradients specify where stratification is stable
against overturn.  In the presence of a phase change, an additional
threshold must obtain for stability due to the thermal effects of the
enthalpy of reaction.  Accordingly, we include a microphysical
inhibition threshold \citep{Markham2022}, based on a critical
``heavy species'' mole-fraction threshold $x_{\rm inhib}$,
\begin{equation}
x_{\rm inhib}
=
\left[
\left(\frac{\mu_{\rm melt} \Delta H}{R T}-1\right)
(\epsilon-1)
\right]^{-1},
\label{eqn:x_inhib}
\end{equation}
where $\mu_{\rm melt}$ is the mean molecular weight of the condensed
phase, $\epsilon \equiv \mu_{\rm vap} / \mu_{\rm H_2}$, where
$\mu_{\rm vap}$ is the mean molecular weight of the heavy condensing
vapor species (e.g., Mg, SiO), and $\Delta H$ is the effective latent
heat per unit mass for condensation.  Physically, $x_{\rm inhib}$
encodes stable stratification in the presence of the phase change.  In
practice, the Ledoux-stable, inhibited region is activated only when
both a stabilizing molecular-weight gradient exists ($\nabla_\mu > 0$)
and the heavy-component abundance exceeds the compositional threshold
($x_{\rm heavy} > x_{\rm inhib}$).  We combine these criteria into a
smooth inhibition ``strength'' $w \in [0,1]$ that controls how strongly
the local temperature gradient is driven toward the radiative
(diffusion-limited) gradient,
\begin{equation}
\begin{aligned}
w &= w_\mu\, w_{x_{\rm inhib}}, \\[0.5ex]
w_\mu &= \mathcal{H}\!\left(\nabla_\mu\right), \\[0.5ex]
w_{x_{\rm inhib}} &= \mathcal{H}\!\left(\frac{x_{\rm heavy}}{x_{\rm inhib}}\right),
\end{aligned}
\label{eqn:w_strength}
\end{equation}
where $\mathcal{H}$ is a Hill function used to smooth the transition
into and out of the inhibited regime and to avoid numerical noise.

The temperature gradient is evaluated by interpolating between a
baseline gradient and the radiative-required gradient,
\begin{equation}
\nabla_{\rm eff}
=
(1-w)\,\nabla
+
w\,\nabla_{\rm rad},
\end{equation}
where $\nabla=\nabla_{\rm ad}$ in convective regions and
$\nabla=\nabla_{\rm rad}$ otherwise.  In the limit $w\rightarrow 1$,
the temperature gradient approaches the diffusion-limited value, while
for $w\rightarrow 0$ the solution reverts to the transport regime
outside the Ledoux-stability region.

Both types of basal boundary layers are global bottlenecks for the
intrinsic luminosity.  When a Ledoux-stable compositional gradient is
present, $L_{\rm int}$ is capped at the maximum luminosity that can be
transmitted through the inhibited layer without violating Ledoux
stability.  This constraint is conceptually analogous to the luminosity
limit imposed by the basal thermal boundary layer of thickness
$\delta_{\rm BL}$, which bounds the net heat flux delivered from the
magma ocean into the envelope.  The distinction between the two lies
in resolvability: the Ledoux-stable region is sufficiently extended
that its stabilizing gradient can be described explicitly, whereas the
basal thermal boundary layer is often too thin to resolve and is
therefore parameterized as a global flux limit.  In the cases of
Uranus and Neptune, the Ledoux-stable region of convective inhibition
is by far the dominant governor on intrinsic luminosity, with the
thin boundary layer reducing the luminosity by factors of order 5 and
the Ledoux region reducing the latter by factors of order 50 times.

\section*{Appendix C: Physical Chemistry} \label{sec:Appendix_C}

Recent work on the physical chemistry of H$_2$--silicate--Fe metal
systems applied to sub-Neptunes \citep{Young_2024,
gilmore_core-envelope_2025, Young2025_Differentiation,
RogersYoungSchlichting2025_MNRAS} shows that the interface separating
outer envelopes and underlying magma oceans is a phase boundary that
can be well approximated as the binodal (or solvus) surface in the
MgSiO$_3$--H$_2$ system \citep{gilmore_core-envelope_2025, Young_2024,
RogersYoungSchlichting2025_MNRAS}.  This boundary moves inward towards
the center as the planet cools
\citep{RogersYoungSchlichting2025_MNRAS}.  Interior to the boundary,
the magma ocean is a supercritical mixture of silicate (modeled as
MgSiO$_3$) and hydrogen with or without a distinct Fe-rich metal phase.
For planets with sufficient hydrogen that their cores have greater
than about $1\%$ by mass H$_2$, Fe metal, MgSiO$_3$ (silicate), and
H$_2$ are entirely miscible \citep{Young2025_Differentiation}.  In
these cases, the cores are a single phase throughout except perhaps at
the most shallow depths.

By determining the common tangents to the Gibbs free energy of mixing
in the MgSiO$_3$--H$_2$ system, using the convex hull for the function
for example, the values for the mole fractions of hydrogen,
$x_{\rm H_2}$, for melt and vapor in equilibrium are obtained.  These
coexisting compositions at various $P$ and $T$ values define the
binodal surface that separates cores from overlying envelopes.  The
non-ideal mixing between MgSiO$_3$ and H$_2$ calculated by
\citet{gilmore_core-envelope_2025} is used here.  Their density
functional theory molecular dynamics (DFT-MD) simulations yield the
binary free energy of mixing expression
\begin{equation}
\begin{split}
\Delta \hat{G}_{\text{binary mix}} &=
\left(L_{\mathrm{CB}} x_{\rm H_2}
+ L_{\mathrm{BC}} (1-x_{\rm H_2})\right)
x_{\rm H_2} (1-x_{\rm H_2}) \\
&\quad \times
\left( 1 - \frac{T}{\tau} + \frac{P}{\pi} \right) \\
&\hspace{-2.0em} + RT \left(
x_{\rm H_2} \ln x_{\rm H_2}
+ (1-x_{\rm H_2}) \ln (1-x_{\rm H_2})
\right),
\end{split}
\label{eqn:MgSiO3_H2}
\end{equation}
where $L_{\mathrm{CB}}$ and $L_{\mathrm{BC}}$ are the binary
interaction parameters and $x_{\rm H_2}$ is the mole fraction of H$_2$
along the binary join.  The values for the subregular mixing
parameters that fit the DFT-MD simulations are
$L_{\mathrm{CB}} = 622000$~J~mol$^{-1}$,
$L_{\mathrm{BC}} = -4950$~J~mol$^{-1}$, $\tau = 4800$~K, and
$\pi = -35$~GPa.  Addition of minor elements like Al has minimal
effects on the position of the binodal surface in
$P$--$T$--$x_{\rm H_2}$ space \citep{gilmore_core-envelope_2025}.

In order to determine the state of the core given the planet's
pressure--temperature structure and its mass fraction of hydrogen we
make use of the ternary phase equilibria that includes iron
\citep{Young2025_Differentiation}.  With this approach, we construct a
thermodynamically consistent ternary mixing model for miscible
MgSiO$_3$--Fe--H$_2$ melts by extrapolating three independently
constrained binary systems (MgSiO$_3$--H$_2$, MgSiO$_3$--Fe, and
Fe--H$_2$) into ternary composition space.  Phase stability is
determined by minimizing the molar Gibbs free energy of mixing at fixed
pressure and temperature.  The total free energy of mixing is written
as the sum of an ideal configurational entropy term and a non-ideal
(excess) mixing term.  The ideal contribution is given by
$RT\sum_i x_i \ln x_i$, where $x_i$ are the mole fractions of Fe (A),
H$_2$ (B), and MgSiO$_3$ (C).  Non-ideal interactions are parameterized
using a subregular solution formalism, with interaction parameters
taken directly from published DFT-MD simulations and experimental
constraints on the three binary joins.

The resulting Gibbs free energy of mixing for the ternary system is
obtained following the Muggianu--Jacob projection method:
\begin{equation}
\begin{aligned}
&\Delta \hat{G}_{\text{mix}} = RT\!\left(
x_{\mathrm{A}}\ln x_{\mathrm{A}} +
x_{\mathrm{B}}\ln x_{\mathrm{B}} +
x_{\mathrm{C}}\ln x_{\mathrm{C}}
\right) \\
& + (1/2) x_{\mathrm{A}} x_{\mathrm{B}}
  \left( L_{\mathrm{AB}} (1 + x_{\mathrm{B}} - x_{\mathrm{A}})
       + L_{\mathrm{BA}} (1 + x_{\mathrm{A}} - x_{\mathrm{B}}) \right) \\
& + (1/2) x_{\mathrm{B}} x_{\mathrm{C}}
  \left( L_{\mathrm{BC}} (1 + x_{\mathrm{C}} - x_{\mathrm{B}})
       + L_{\mathrm{CB}} (1 + x_{\mathrm{B}} - x_{\mathrm{C}}) \right) \\
& \quad \times (1 - T/\tau + P/\pi) \\
& + (1/2) x_{\mathrm{C}} x_{\mathrm{A}}
  \left( L_{\mathrm{CA}} (1 + x_{\mathrm{A}} - x_{\mathrm{C}})
       + L_{\mathrm{AC}} (1 + x_{\mathrm{C}} - x_{\mathrm{A}}) \right)
\end{aligned}
\label{eqn:Gmix}
\end{equation}
where $x_{\mathrm{A}}$, $x_{\mathrm{B}}$, and $x_{\mathrm{C}}$ denote
the mole fractions of Fe, H$_2$, and MgSiO$_3$, respectively, with
$x_{\mathrm{A}}+x_{\mathrm{B}}+x_{\mathrm{C}}=1$.  The MgSiO$_3$--H$_2$
interaction parameters are
$L_{\mathrm{CB}}=6.22\times10^{5}$~J~mol$^{-1}$ and
$L_{\mathrm{BC}}=-4.95\times10^{3}$~J~mol$^{-1}$, with temperature and
pressure dependence governed by $\tau=4800$~K and $\pi=-35$~GPa
\citep{gilmore_core-envelope_2025}.  The MgSiO$_3$--Fe interaction is
treated as a regular solution with
$L_{\mathrm{AC}}=L_{\mathrm{CA}}=2.4\times10^{5}-28T+1116P$
(J~mol$^{-1}$), based on DFT-MD calculations for MgO--Fe mixing
\citep{Young2025_Differentiation}.  The Fe--H$_2$ interaction
parameters are $L_{\mathrm{AB}}=1.38\times10^{5}-9500P$ and
$L_{\mathrm{BA}}=1.7\times10^{4}-9500P$ (J~mol$^{-1}$), calibrated to
reproduce experimental and \emph{ab initio} constraints on hydrogen
solubility in molten iron.  No explicit, and unknown, ternary
interaction parameter is included ($L_{\mathrm{ABC}}=0$), as is often
the case in such applications.

With this free energy of mixing we can map the regions of miscibility
and immiscibility between MgSiO$_3$, Fe, and H$_2$ as a function of
pressure and temperature.

\section*{Appendix D: Material Properties} \label{sec:Appendix_D}

For liquid silicate densities, we use a previously derived approach
\citep{Young2025_Differentiation}.  Briefly, for the silicate melt we
use an equation of state fit to the MgSiO$_3$ liquid properties
\citep{DeKoker2009} using the algorithms of \citet{Wolf_2018}.  Total
pressure is computed by combining the elastic (cold) compression term
from a Vinet equation of state with thermal pressure contributions
\citep{Wolf_2018}.  For liquid iron, we use a Vinet equation of state
modified to include thermal pressure from \citet{Kuwayama2020}.  The
Gr\"uneisen parameter is a simple function of the compression ratio,
which in turn determines the thermal contribution to pressure.

We find that addition of H (and by fiat, Fe) to the supercritical magma
ocean changes its density.  DFT-MD simulations show that the presence of
hydrogen in the silicate melt reduces its density such that a linear
mixture of the compressed densities of MgSiO$_3$ and H$_2$ by volume
reproduces the density of the mixture at the conditions of the binodal
\citep{Young2025_Differentiation}.  We therefore include the effects
of H$_2$ on the density of the supercritical melt by calculating the
density of the mixture at the pressures and temperatures of the
surface of the magma ocean, $\rho_0$.  The density of the mixture is
\begin{equation}
\frac{1}{\hat{V}_{\rm mix}} =
    x_{\text{H}_2}\frac{1}{\hat{v}_{\text{H}_2}}
  + x_{\text{sil}}  \frac{1}{\hat{v}_{\text{sil}}},
\label{eq:Vmix}
\end{equation}
where $\hat{v}_{\text{H}_2}$ and $\hat{v}_{\text{sil}}$ are the molar
volumes, and where $x_{\text{H}_2}$ and $x_{\text{sil}}$ are the mole
fractions of hydrogen and silicate in the binary mixture.  Densities
are obtained from $\mathrm{MW}_{\rm mix}/\hat{V}_{\rm mix}$, where
$\mathrm{MW}_{\rm mix}$ is the molecular weight of the mixture.  Molar
volumes are obtained from densities and the equations of state using
$\rho_i/\mathrm{MW}_i$.  We fixed the densities to be 0.09~g~cm$^{-3}$
and 2.5~g~cm$^{-3}$ for H$_2$ and silicate, respectively, at the
surface of the magma ocean.  These values come from the equations of
state for hydrogen \citep{Chabrier2019} and \emph{ab initio} molecular
dynamics simulations of MgSiO$_3$ melt at 6000~K and 3.5~GPa
\citep{Young2025_Differentiation}.

\begin{figure}
\centering
\includegraphics[width=0.97\columnwidth]{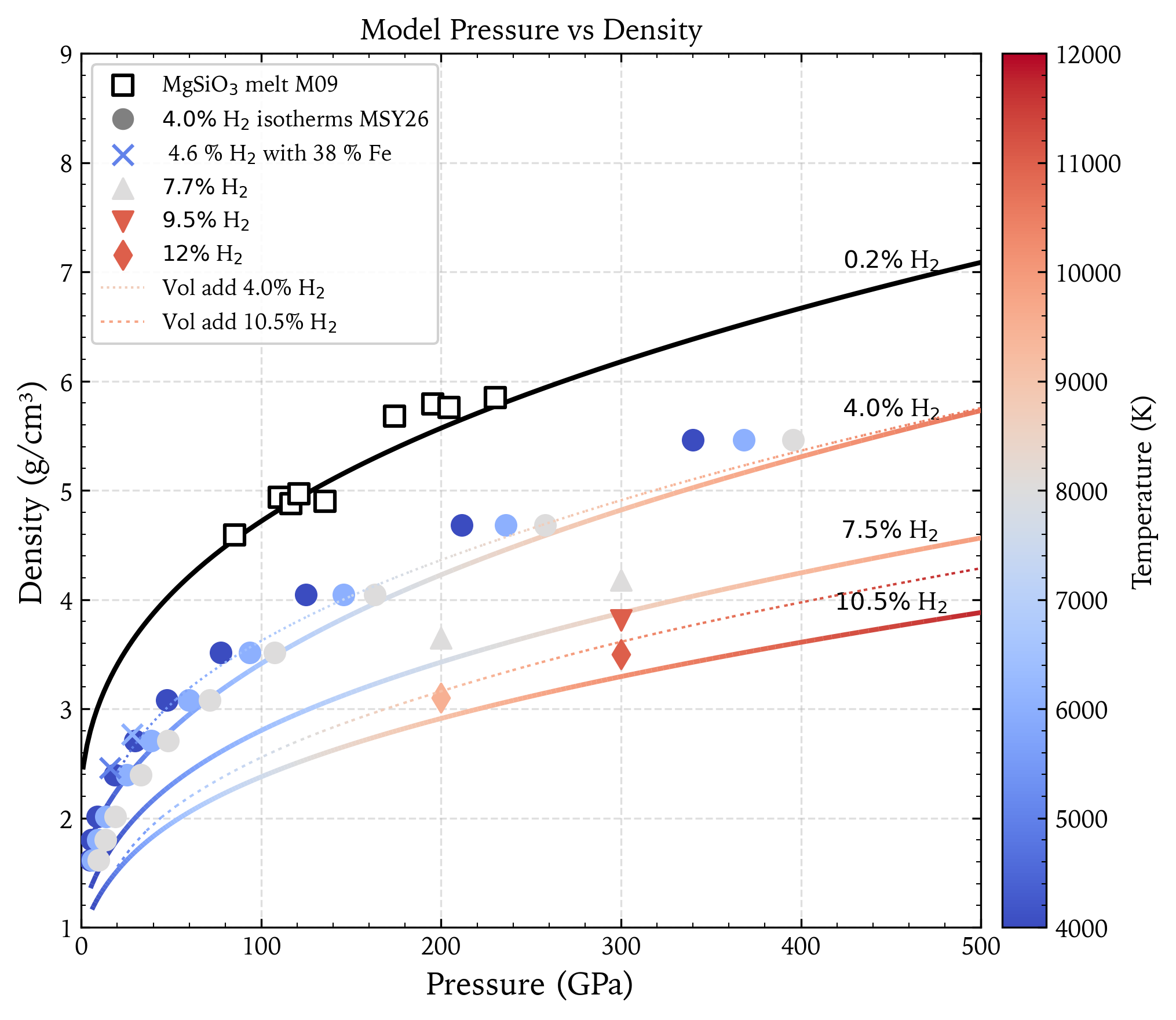}
\caption{Equation of state and density data for MgSiO$_3$ and supercritical mixtures of MgSiO$_3$ and H$_2$ from
DFT-MD simulations (circles and diamonds) compared with experiments (squares, \citealt{Mosenfelder_2009}, M09) and model adiabats using our EoS for this study. Curves for simple volume-additive densities are shown for comparison. Adiabat curves using the EoS used here are labeled with the H$_2$ concentrations in weight percent for the miscible silicate-H$_2$ mixtures. Corresponding volume-additive curves are shown for the $4\%$ and $10.5\%$ cases. Note adiabats are shallower than isotherms on this diagram. }
\label{fig:EoS}
\end{figure}

Deeper in the supercritical magma ocean, the effects of hydrogen on the
compressibilities must be accounted for. Based on a complete equation of state
for the mixture of MgSiO$_3$ and $4\%$ by weight H$_2$ \citep{Marcum2026} we
determined that the reference bulk modulus, $K_T$, is similar for pure MgSiO$_3$ liquid and for the supercritical mixture containing $4\%$ H$_2$. However, the pressure derivative of the bulk modulus, $K'$, decreases from 5.71 for  MgSiO$_3$ liquid to 4.32 for the H-bearing mixture \citep{Marcum2026}. The smaller value of $K'$ indicates that the supercritical mixture stiffens less rapidly with pressure, so that added hydrogen both lowers the density and increases the compressibility of the phase at high pressure. For H-bearing compositions, we therefore approximate the pressure derivative of the bulk modulus by linear interpolation/extrapolation in H$_2$ mass fraction:
\begin{equation}
K'(w_{\rm H_2}) =
K'_{\rm pure}
+ \left(K'_{4\%\,{\rm H}_2} - K'_{\rm pure}\right)
\left(\frac{w_{\rm H_2}}{0.04}\right).
\label{eq:kprime}
\end{equation}
Here $K'_{\rm pure}$ denotes pure MgSiO$_3$ liquid, with a value of $7.42$, $K'_{4\%\,{\rm H}_2}$ denotes the supercritical mixture with $4\%$ H$_2$ by mass, with a value of $4.47$, and $w_{\rm H_2}$ is the H$_2$ mass fraction.  This approximation is tested
against densities from additional DFT-MD simulations at conditions near to those obtained for the interiors of Neptune and Uranus using the methods described previously \citep{Young2025_Differentiation, Marcum2026}.  We find that using Equation \ref{eq:kprime}, we reproduce the densities at similar temperatures and pressures, albeit with some differences in the weight fractions of hydrogen in the melt.

A comparison between experimental data for MgSiO$_3$, the complete equation of state for the mixture with
$4\%$ H$_2$, and the extrapolation to higher concentrations of hydrogen up to $12\%$ by mass is shown in Figure \ref{fig:EoS}. The plot illustrates that our densities are broadly similar to those predicted by simple volume-additive mixing of densities for pure MgSiO$_3$ and pure H$_2$ using their respective equations of state, except at the lowest pressures (compare dotted and solid curves in Figure \ref{fig:EoS}). In greater detail, however, one can also see that for the $10.5\%$ H$_2$ adiabat, for example, similar to our Uranus model, the DFT-MD simulations produce densities similar to our extrapolated values but with $12\%$ H$_2$ rather than $10.5\%$ H$_2$. Accounting for this discrepancy would require the same pressure-density curves to be calculated using greater total mass fractions of H$_2$ for our models.  We observe that changing the parameters of the EoS has this effect; the same pressure-density curves fit the planets but with higher total mass fractions of hydrogen. The higher mass fractions of hydrogen are associated with lower temperatures along the adiabats.  Until the full EoS of this supercritical melt phase as a function of H$_2$ concentration is firmly established, we conclude that going beyond the linear extrapolation of $K'$ is neither warranted nor justified, and accept that there is a systematic uncertainty in the absolute magnitude of the total hydrogen mass fraction in our models.   

We model Fe dissolved in a supercritical silicate--H$_2$ melt by
assigning Fe a partial molar volume that reflects Fe occupation of
Mg-like cation sites within the silicate melt framework
\citep{Young2025_Differentiation}.  The Fe-site partial molar volume
is
\begin{equation}
\bar V_{\mathrm{Fe\, site}}^{\mathrm{eff}}
= \phi\,\alpha\;\bar V_{\mathrm{MgSiO_3}}^{\mathrm{pure}}(P,T),
\end{equation}
and the corresponding effective partial Fe density is
\begin{equation}
\rho_{\mathrm{Fe\, site}}^{\mathrm{eff}}
= \frac{\mathrm{MW}_{\mathrm{Fe}}}{\bar V_{\mathrm{Fe\, site}}^{\mathrm{eff}}},
\end{equation}
where $\phi$ denotes the fraction of the pure MgSiO$_3$ formula-unit
molar volume attributable to the Mg site,
$\alpha=(r_{\mathrm{Fe}}/r_{\mathrm{Mg}})^3$ scales the site volume by
the ratio of octahedrally coordinated ionic radii
\citep{Shannon_1976}, and
$\bar V_{\mathrm{MgSiO_3}}^{\mathrm{pure}}=
 \mathrm{MW}_{\mathrm{MgSiO_3}}/\rho_{\mathrm{MgSiO_3}}^{\mathrm{pure}}$
is the molar volume of pure MgSiO$_3$ at the local $P$ and $T$.
Assuming additivity of specific volumes (no excess mixing volume), the
density of the miscible melt with Fe mass fraction $w_{\mathrm{Fe}}$
is given by
\begin{equation}
\frac{1}{\rho_{\mathrm{mix}}}
= \frac{1-w_{\mathrm{Fe}}}{\rho_{\mathrm{MgSiO_3+H_2}}}
+ \frac{w_{\mathrm{Fe}}}{\rho_{\mathrm{Fe}}^{\mathrm{eff}}},
\end{equation}
where $\rho_{\mathrm{MgSiO_3+H_2}}$ is the density of the
MgSiO$_3$+H$_2$ mixture derived at $P$ and $T$.  An empirical value
for $\phi$ of 0.5 is obtained from the MgO--Fe system
\citep{Insixiengmay_2025}. The effect of Fe on the supercritical mixture as determined using DFT-MD with methods similar to \cite{Insixiengmay_2025} is
illustrated by the two crosses in Figure \ref{fig:EoS} where it is seen to be relatively minor. 

\begin{figure}
\centering
\includegraphics[width=0.95\columnwidth]{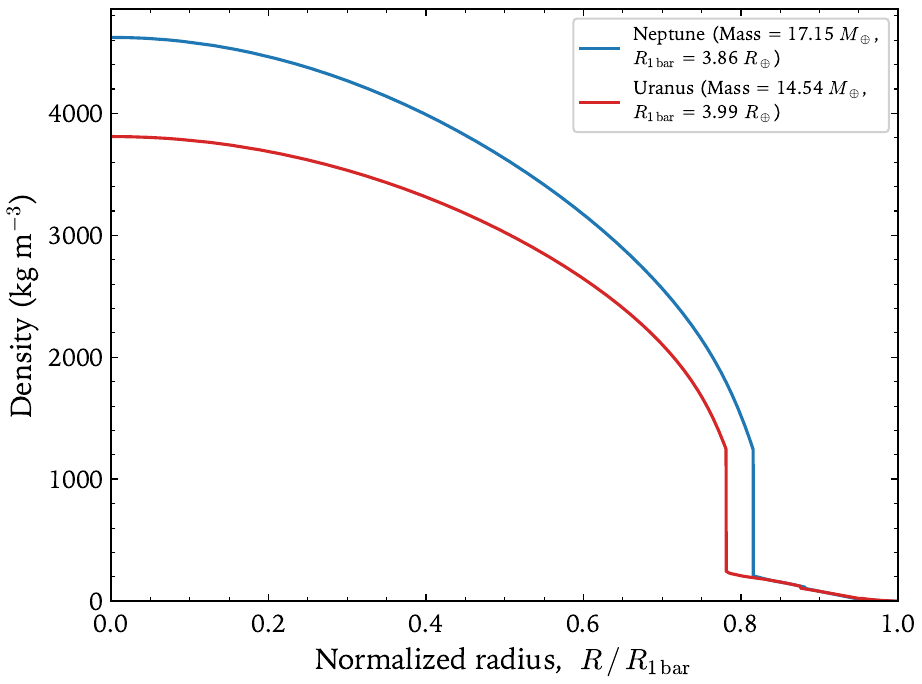}
\caption{Density versus normalized radial position for the Neptune and Uranus best-fit models. }
\label{fig:rho_profiles}
\end{figure}

The densities versus normalized planet radii for both Neptune and Uranus based on the equations of state described above are shown in Figure \ref{fig:rho_profiles}. 

We estimate the depression of the MgSiO$_3$ solidus due to dissolved
H$_2$ using the cryoscopic equation for an ideal solution:
\begin{equation}
\Delta T_m = -\frac{R\,T_m^2}{\Delta H_{\rm fus}}\,\ln X_{\rm MgSiO_3},
\label{eqn:cryo}
\end{equation}
where $T_m$ is the pure-component melting temperature,
$\Delta H_{\rm fus}$ is the molar enthalpy of fusion, and
$X_{\rm MgSiO_3}$ is the mole fraction of MgSiO$_3$ in the melt.  At
a binodal condition of $P_{\rm binodal} \approx 5$~GPa
and $T \approx 3300$~K, the model H$_2$ mass fraction in the
silicate melt is $f_{\rm H_2}^{\rm core} \approx 3$~wt\%,
corresponding to $X_{\rm MgSiO_3} = 0.38$.  Taking $T_m = 3000$~K
and $\Delta H_{\rm fus} = 75$~kJ~mol$^{-1}$ at 5~GPa
\citep{Stixrude2005, Stixrude2014}, the ideal-mixing estimate yields
$\Delta T_m \approx 965$~K, depressing the effective MgSiO$_3$
solidus from $\sim 3000$~K to $\sim 2035$~K and placing the binodal
conditions well within the melt field.  The enthalpy of fusion carries
substantial uncertainty (published \emph{ab initio} values span
50--200~kJ~mol$^{-1}$), but even at the upper bound the depression
exceeds several hundred degrees, making this conclusion reasonably robust.

The melting point depression at fixed composition in
Equation~(\ref{eqn:cryo}) depends on pressure through $T_m(P)$ and
$\Delta H_{\rm fus}(P)$.  The MgSiO$_3$ Clausius--Clapeyron slope is
large and positive \citep{Stixrude2005}, with $T_m$ rising from
$\sim 3000$~K at 5~GPa to $\sim 7000$--$8000$~K at 300~GPa.  Since
$\Delta T_m \propto T_m^2 / \Delta H_{\rm fus}$, two competing effects
operate with increasing pressure: $T_m^2$ grows rapidly, amplifying the
depression, while $\Delta H_{\rm fus}$ also increases moderately as the
liquid--solid density contrast diminishes \citep{Stixrude2005}.  The
$T_m^2$ term dominates, so the ratio
$T_m^2 / \Delta H_{\rm fus}$ grows with pressure and the cryoscopic
depression is expected to increase throughout the interior.  Taking
$T_m \approx 7500$~K and $\Delta H_{\rm fus} \approx 150$~kJ~mol$^{-1}$
at 300~GPa as rough \emph{ab initio} extrapolations, the ideal-mixing
estimate yields $\Delta T_m \approx 3016$~K.  This estimate is illustrative only given the uncertainties in
$\Delta H_{\rm fus}$ and H$_2$ solubility at these extreme conditions,
but suggests that interiors modeled here should be entirely molten.

Iron mixing and non-ideal H$_2$--MgSiO$_3$ mixing should further lower
the melting point.  For example, the activity coefficient
$\gamma_{\rm MgSiO_3} = 0.42$ from the regular-solution model used in
our study \citep{gilmore_core-envelope_2025}.  Replacing
$X_{\rm MgSiO_3}$ with the activity
$\gamma_{\rm MgSiO_3} X_{\rm MgSiO_3} = 0.16$ in
Equation~(\ref{eqn:cryo}), the estimated depression in melting
temperature is $\sim 1830$~K, indicating that non-ideality acts in the
direction of stabilizing the melt phase.

\section*{Appendix E: Winds \label{sec:Appendix_E}}
\begin{figure*}[t!]
   \centering
   \includegraphics[width=0.85\textwidth]{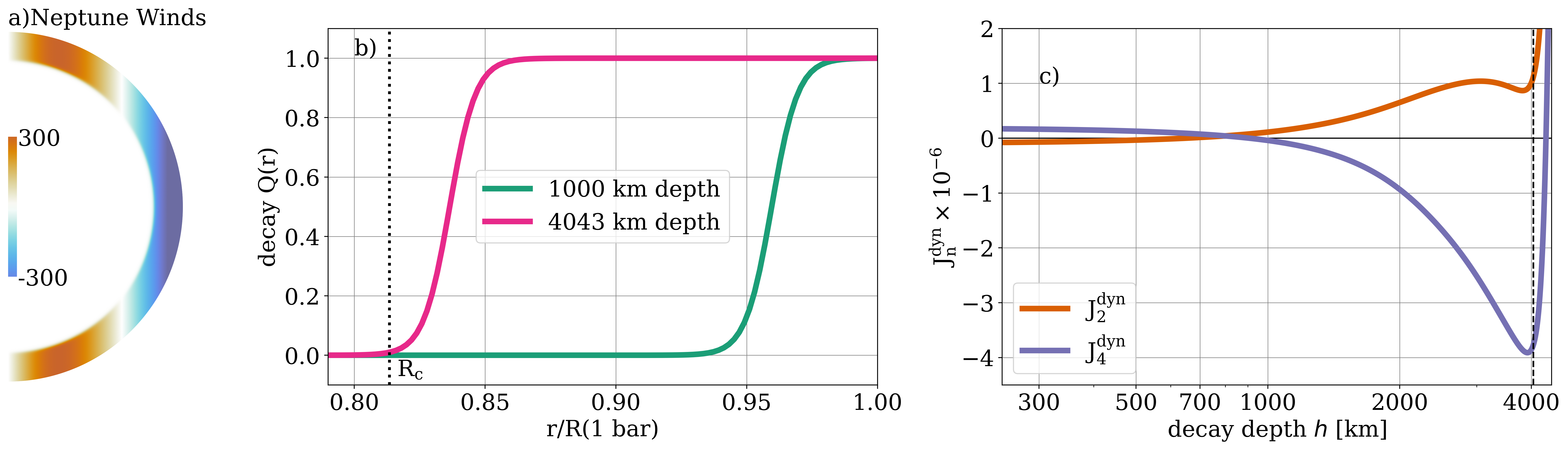}
   \includegraphics[width=0.85\textwidth]{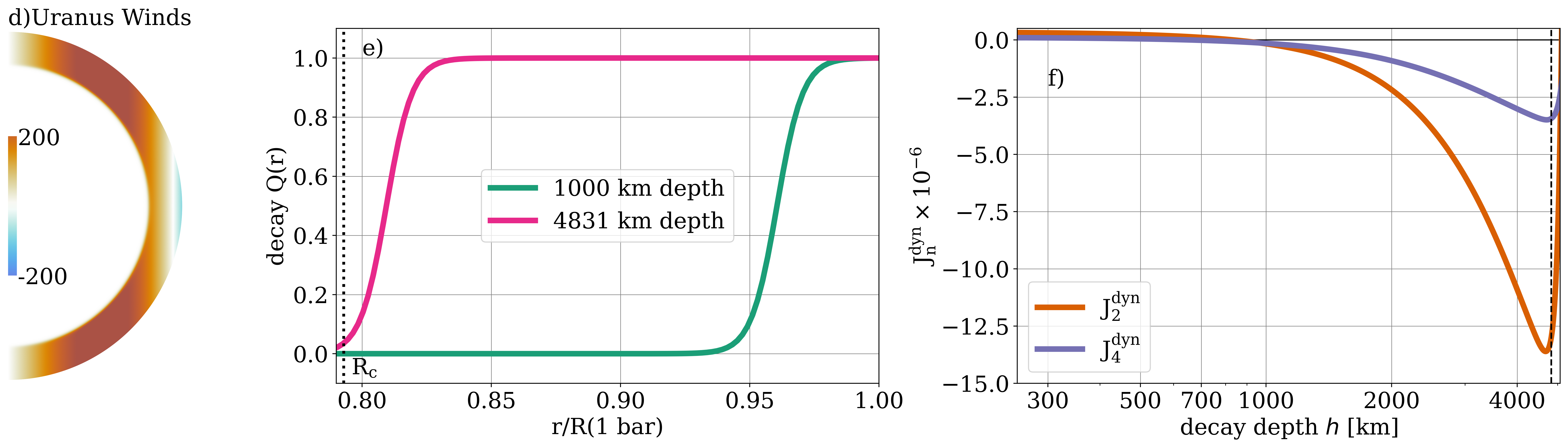}
      \caption{Plots illustrating wind models.  Panels a) and d) show meridional cuts of the surface-measured zonal winds extrapolated downward - aligned with the axis of rotation - and reaching the bottom of the convective atmosphere of Neptune and Uranus, respectively. Red/blue indicates prograde/retrograde flow in m/s. Panels b) and e) show the radial decay profile of the winds $Q(r)$, with radius on the x-axis normalized with respect to the 1-bar equatorial radius of the model. We show both our assumed, deep, profiles, as well as profiles reaching 1,000~km, similar to those inferred by \citet{Kaspi2013}. The  dotted vertical line indicates the location of the core radius. Panels c) and f) are the values of $J_2^{dyn}$ and $J_4^{dyn}$ as a function of wind depth, where the dashed vertical line indicates 3838~km and 4584~km depth, respectively.
      }
         \label{fig:Winds}
\end{figure*}

Our approach in calculating the dynamic components of the gravity signature follows that outlined in \citet{Wicht_2020, Dietrich_2021} and is adapted to the even gravity harmonics available for the ice giants as in \citet{Soyuer_2023}. We refer the reader to these works for further details and only outline the specifics of how the method was applied here.

The zonal gravity harmonic components due to the winds are defined as
\begin{equation}
    J_n^{\rm dyn}=-\frac{1}{MR^n}\int_V\rho^{\rm dyn}(\boldsymbol{r})P_n(\cos\theta)r^n\, dV,
\end{equation}
having expanded both the density and gravity in terms of a static (background state) and dynamic (due to the winds) component:
\begin{equation}
    \rho = \rho^{\rm stat}+\rho^{\rm dyn},\quad J_n^{\rm obs}=J^{\rm stat}_n+J_n^{\rm dyn}.
\end{equation}
Thus, to obtain $J_n^{\rm dyn}$, we must link the zonal flows with their associated density perturbations. For this, we follow exactly the steps detailed in \citet{Dietrich_2021}. This includes the dynamic self-gravity \citep{Wicht_2020} and omits the centrifugal potential \citep{Cao_2017, Galanti_2017}.

We use the analytical forms of the surface wind profiles, as a function of the colatitude $\theta$, which are the fits to the observed cloud-tracking measurements. For Neptune this is given by
\begin{equation}
    U_N(\theta)=\sum^6_{i=1} c_i\sin((2i-1)\theta),
\end{equation}
from \citet{Soyuer_2023}, while for Uranus we have
\begin{equation}
    U_U(\theta)=170\times(0.6\sin\theta+\sin3\theta)\;\rm m/s,
\end{equation}
from \citet{Hammel_2001}. These are extrapolated downwards, with the surface structures taken to be invariant with respect to the axis of rotation (see Fig~\ref{fig:Winds}a and d). We then assume a radially-dependent decay profile of the winds, with the functional form
\begin{equation}
    Q(r) = \frac{\tanh((r-(1-h))/\delta_h) +1}{\tanh(h/\delta_h) +1},
\end{equation}
where we fix the width of the hyperbolic tangent $\delta_h=0.01$ and vary the depth $h$ (see Fig.~\ref{fig:Winds}b and e).

As pointed out by \citet{Cao_2017}, to reliably constrain the wind structure with only even gravity harmonics, the contributions beyond $J_6$ are required. For the lower harmonics, small uncertainties in the background profiles match the amplitudes of the dynamic contributions. Therefore, we do not take an optimization approach to the evaluation of the wind depths, and only iterate once between the calculation of $J^{\rm dyn}$ and adjustment of the static modeling. We show the calculated values of $J_2^{dyn}$ and $J_4^{dyn}$ for both planets as a function of zonal wind depth in Fig.~\ref{fig:Winds}c and f (from model $\rho^{\rm stat}$ and $g^{\rm stat}$ as inputs). 

\begin{table}[htbp]
\centering
\caption{Dynamic gravity moments based on winds extending to near the bottom of each planet's convective atmosphere.}
\label{tab:DynJ}
\begin{tabular*}{\columnwidth}{@{\extracolsep{\fill}}lccc@{}}
\hline\hline
Planet & Depth [km] & $J_2^{\rm dyn}\,(\times10^{-6})$ & $J_4^{\rm dyn}\,(\times10^{-6})$ \\
\hline
Neptune & 3838 & 0.87 & -3.88 \\
Uranus & 4584 & -13.47 & -3.46 \\
\hline\hline
\end{tabular*}
\end{table}

For our reference values of $J_2^{dyn}$ and $J_4^{dyn}$ we take those corresponding to a wind depth based on the bottom of the convecting atmosphere in a previous iteration of the static models. These are given in Table~\ref{tab:DynJ}.

\begin{acknowledgments}
We acknowledge financial support from NASA grant 80NSSC24K0544
(Emerging Worlds program). P.N.W. acknowledges support from the NASA Juno project.
\end{acknowledgments}

\newpage

\bibliographystyle{aasjournal}
\bibliography{edy_references}

\end{document}